\documentclass[%
 reprint,
 amsmath,amssymb,
 aps,
]{revtex4-2}

\usepackage{graphicx}
\usepackage{dcolumn}
\usepackage{bm}
\usepackage{color}
\usepackage{textgreek}
\usepackage{chemformula}
\usepackage[hidelinks]{hyperref}

\begin{document}

\preprint{APS/123-QED}

\title{Electron-positron pairs and radioactive nuclei production by irradiation of high-Z target with \textgamma-photon flash generated by an ultra-intense laser in the $\lambda^3$ regime}

\author{David Kolenatý} 
 \altaffiliation[Also at ]{Department of Physics and NTIS – European Centre of Excellence, University of West Bohemia, Univerzitní 8, 306 14 Plzeň, Czech Republic.}
\email{kolenaty@kfy.zcu.cz}
\author{Prokopis Hadjisolomou}%
\author{Roberto Versaci}
\author{Tae Moon Jeong}
\author{Petr Valenta}
 \altaffiliation[Also at ]{Faculty of Nuclear Sciences and Physical Engineering, Czech Technical University in Prague, Břehová 7, Prague 11519, Czech Republic.}
\author{Veronika Olšovcová}
\author{Sergei Vladimirovich Bulanov}
 \altaffiliation[Also at ]{National Institutes for Quantum Science and Technology (QST), Kansai Photon Science Institute, 8-1-7 Umemidai, Kizugawa, Kyoto 619-0215, Japan.}
\affiliation{ELI Beamlines Centre, Institute of Physics, Czech Academy of Sciences, Za Radnicí 835, 25241 Dolní Břežany, Czech Republic}

\date{\today}

\begin{abstract}

\par This paper studies the interaction of laser-driven \textgamma-photons and high energy charged particles with high-Z targets through Monte-Carlo simulations. The interacting particles are taken from particle-in-cell simulations of the interaction of a tightly-focused ultraintense laser pulse with a titanium target. Lead is chosen as the secondary high-Z target owing to its high cross section of the giant dipole resonance and electron-positron pair production. The results reveal an ultra-short ultra-relativistic collimated positron population and their energy spectra, angular distribution, and temporal profile are found. We investigate the target thickness dependence of the resulting total numbers and total kinetic energies of various particle species emitted from the lead target irradiated with laser-generated \textgamma-photons and charged particles separately. We plot the charts of residual high-Z nuclides generated by irradiation of the lead target. Owing to the short pulse duration, the \textgamma-photon, electron-positron, and neutron sources can find applications in material science, nuclear physics, laboratory astrophysics, and as injectors in laser-based accelerators of charged particles. 

\end{abstract}

\maketitle


\section{\label{sec:level1}Introduction}

\par The invention and development of the Chirped Pulse Amplification (CPA) technique \cite{1985_StricklandD} coupled with the Kerr lens mode-locking technique \cite{1991_SpenceDE} led to the rapid growth of laser power during the last few decades \cite{2019_DansonCN}. The record power of a 10 PW laser has been recently announced by ELI-NP \cite{2020_TanakaKA} and another 10 PW laser is being finalized at ELI-Beamlines. Current worldwide activities on PW laser systems and plans to construct lasers beyond 100 PW are outlined in this review article \cite{2019_DansonCN}. Intensities as high as $10^{23} \text{ W/cm}^{2}$ have been recently reported \cite{2021_YoonJW}. The highest intensity achievable by a laser of certain energy corresponds to a tightly-focused single-cycle laser pulse as proposed by \cite{2002_MourouG}, referred to as the $\lambda^{3}$ regime, where $\lambda$ is the laser wavelength.

\par The laser-matter interactions at such ultrahigh intensities result not only in acceleration of electrons \cite{2009_EsareyE} and ions \cite{2012_DaidoH} but also in the generation of high-energy \textgamma-photons and electron-positron pairs \cite{2006_MourouG}. As theoretically foreseen \cite{2012_NakamuraT,2012_RidgersCP} and recently calculated by three-dimensional (3D) Particle-In-Cell (PIC) simulations \cite{2018_LezhninKV,2021_HadjisolomouP,2022_hadjisolomouP}, the laser-matter interactions at ultrahigh intensities result in high conversion efficiency of the laser energy to \textgamma-photon energy, up to 50\%  in a time comparable to the laser pulse duration. The generation of such \textgamma-photon flash is one of the main goals of multi-PW laser facilities \cite{2016_MckennaP}. 

\par All aforementioned laser-generated high-energy particles undergo many interactions with the surrounding matter \cite{2009_PerkinsDH}. They generate further electrons through ionization \cite{1944_LandauL} but also electron-positron pairs through pair production in the Coulomb field of nuclei \cite{1934_BetheH} and atomic electrons \cite{1939_WheelerJA}. Additionally, \textgamma-photons are produced by Bremsstrahlung \cite{1959_KochHW} of electrons, positrons, and ions or come from nuclear interactions of heavy ions \cite{1991_AichelinJ}. All \textgamma-photons undergo Compton scattering \cite{1923_ComptonAH} on electrons. Furthermore, \textgamma-photons, neutrons, protons, ions, and radioactive nuclides are produced through photonuclear \cite{1970_HaywardE,2004_NedorezovVG,2021_NedorezovVG} and electronuclear \cite{1975_BudnevVM} reactions, and through nuclear interactions of heavy ions \cite{1991_AichelinJ}. All these interactions and properties of generated particles can be simulated by Monte Carlo (MC) particle transport codes.

\par The pulsed sources of particles with such high energies can be used in the field of fundamental science as a positron injector for plasma-based wakefield accelerators \cite{2019_AlejoA}, for electron-positron pair plasma studies of astrophysical phenomena \cite{2010_ChenH, 2011_ChenH,2015_ChenH,2015_SarriG}, but also can find applications in medicine and material science as positron annihilation lifetime spectroscopy \cite{2016_XuT,2021_AudetTL,2021_SarriG}, neutron resonance spectroscopy \cite{2010_HigginsonDP, 2020_ZimmerM}, neutron diffraction \cite{2016_AksenovVL,2020_VogelS}, and nuclear waste management \cite{2001_RevolJP, 2008_LensaW}. Activation of residual nuclides can be used for the study of nuclear physics and astrophysics \cite{2016_NishiuchiM}, and for direct applications in nuclear medicine \cite{2021_SunZ}.

\par In this work we couple a 3D PIC simulation with MC simulations to examine the interaction of the \textgamma-photon flash and high energy charged particles generated by an ultra-intense laser in the $\lambda^3$ regime with a high-Z target, specifically lead. We consider the possible interactions of these laser-driven high energy particles with a lead cylindrical target of varying thickness in order to estimate the activation of produced residual nuclei and the properties of particles emitted from the target, with the focus on positrons in view of the aforementioned applications.

\par This paper is organized as follows. In Section~\ref{sec:level2}, we describe the simulation setup of the PIC and the MC simulations and the result of the PIC simulation of laser-matter interaction. In Section~\ref{sec:level3}, we show and discuss the results of the MC simulations in terms of generated nuclides and emitted particles dependence on the thickness of the lead target, the energy and angular distributions of positrons emitted from the target with various thicknesses, the temporal profile of collimated emission of ultra-relativistic positrons, and the distribution of generated residual nuclides. In Section~\ref{sec:level4}, we conclude our findings.


\section{\label{sec:level2}Simulation Setup}

\par
\begin{figure}[h]
\includegraphics[width=.47\textwidth]{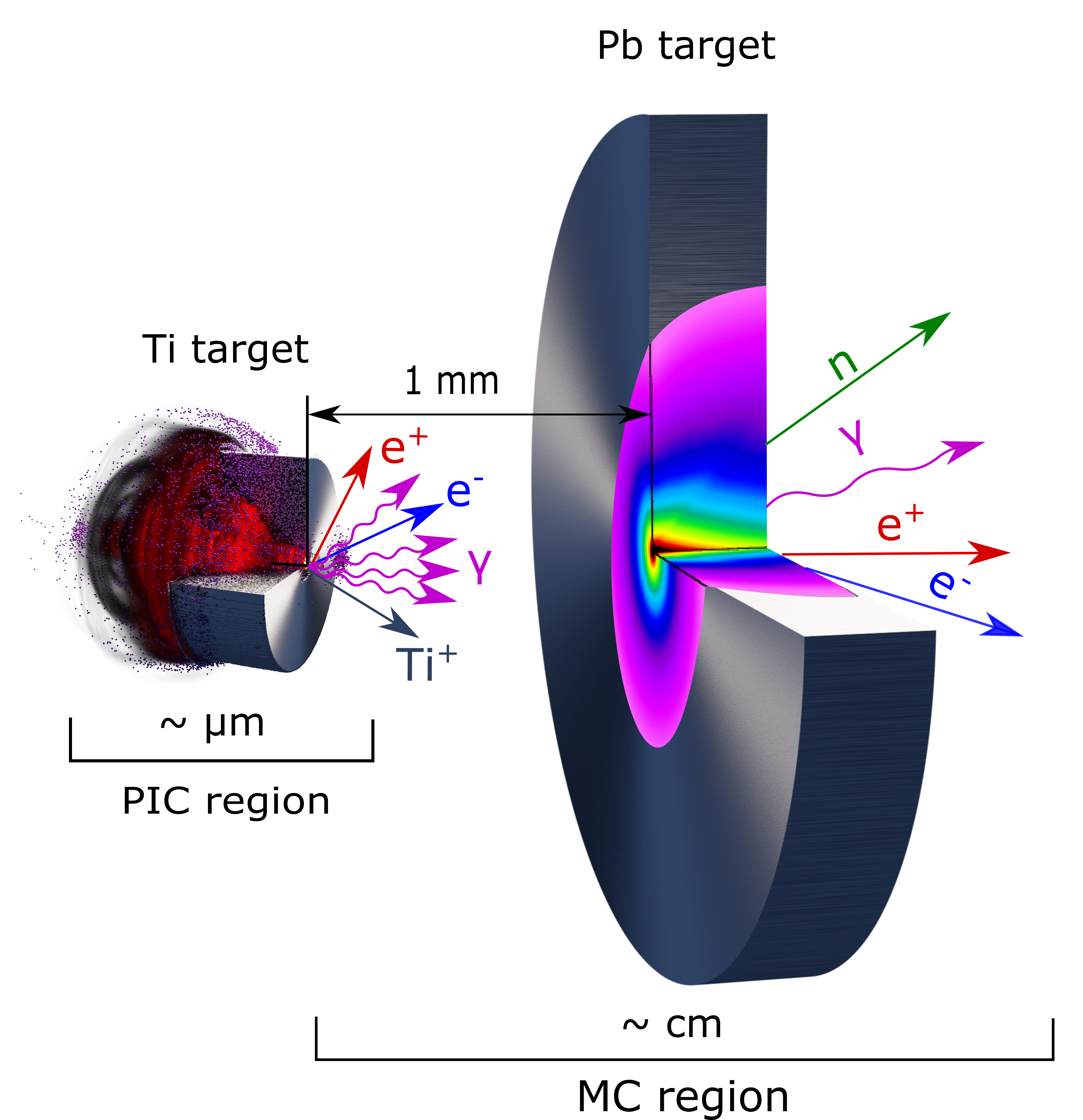}
\caption{\label{fig1} The drawing illustrates the geometrical setup and the scales of PIC and MC simulation regions. The laser-matter interaction resulting in the generation of high energy particles is computed with the PIC code using a titanium target. The simulation employs the $\lambda^{3}$ regime, where a radially polarized near-single-cycle pulse is focused to a spherical-like intensity of $\sim 0.5\lambda$ FWHM spatial profile and $\sim3.4 \text{ fs}$ pulse duration. An $\sim80 \text{ PW}$ power laser under the tight focusing scheme used reached an intensity of $10^{25} \text{ Wcm}^{-2}$. The interactions of laser-driven high energy charged particles and \textgamma-photons with high-Z matter and the resulting products were computed with a MC code using a lead target. The deposited energy to the lead target is represented by color mapping.}
\end{figure}

 The output from the PIC simulation of ultra-intense laser pulse interaction with a titanium target is imported into MC simulations as the primary particles (\textgamma-photons, electrons, positrons, and ions) to compute their interactions with a lead target and the resulting products. The geometrical setup of the MC simulations in respect to the PIC coordinates is designed to irradiate the lead target with almost all of the particles moving in the forward direction, i.e., with a positive momentum component in the laser propagation direction, see Fig.~\ref{fig1}. The titanium target for the PIC simulation is chosen since its electron number density is optimum for high conversion efficiency of the laser energy to \textgamma-photon energy \cite{2021_HadjisolomouP}. Natural lead ($Z = 82$, $A = 207.19$) is chosen as target material for the MC simulations owing to high cross sections for electron-positron pair production, proportional to $Z^{2}$, and for the GDR being the dominant photonuclear interaction for the energies considered. The detailed descriptions of the simulations are given in the following subsections.

\subsection{\label{sec:level2a}Particle in Cell Simulation}

\par This subsection provides a short description of the PIC simulation presented in Ref.~[\onlinecite{2021_HadjisolomouP}]. The simulation is performed by the 3D EPOCH \cite{2015_ArberTD} PIC code. The simulation employs the $\lambda^{3}$ regime \cite{2002_MourouG}, where a radially polarized near-single-cycle pulse is focused by an $f$-number $1/3$ parabola \cite{2015_JeongTM,2018_JeongTM} to a spherical-like intensity of $\sim 0.5\lambda$ FWHM spatial profile and $\sim3.4 \text{ fs}$ pulse duration. An $\sim80 \text{ PW}$ power laser under the tight focusing scheme used reached an intensity of $10^{25} \text{ Wcm}^{-2}$.

\par The radius of the target, $r = 2.4 \kern1ex \mathrm{\mu m}$, is large enough for the target destruction by the laser not to reach the target edges. The highest conversion efficiency of the laser energy to \textgamma-photon energy corresponds to a relatively thick target ($2 \kern1ex \mathrm{\mu m}$) and an electron number density of $\sim1.25 \times 10^{24} \text{ cm}^{-3}$, similar to that of titanium. The cubic simulation box is split into $512^{3}$ cubic cells with an edge size of 20 nm containing 8 macroparticles each. The size of the simulation box of $10.24 \kern1ex \mathrm{\mu m}$ and the 16 fs duration of the simulation are both large enough for the conversion of laser energy to the energy of all species to saturate before the fields or particles exit the simulation box.

\par
\begin{figure}[h]
\includegraphics[width=.47\textwidth]{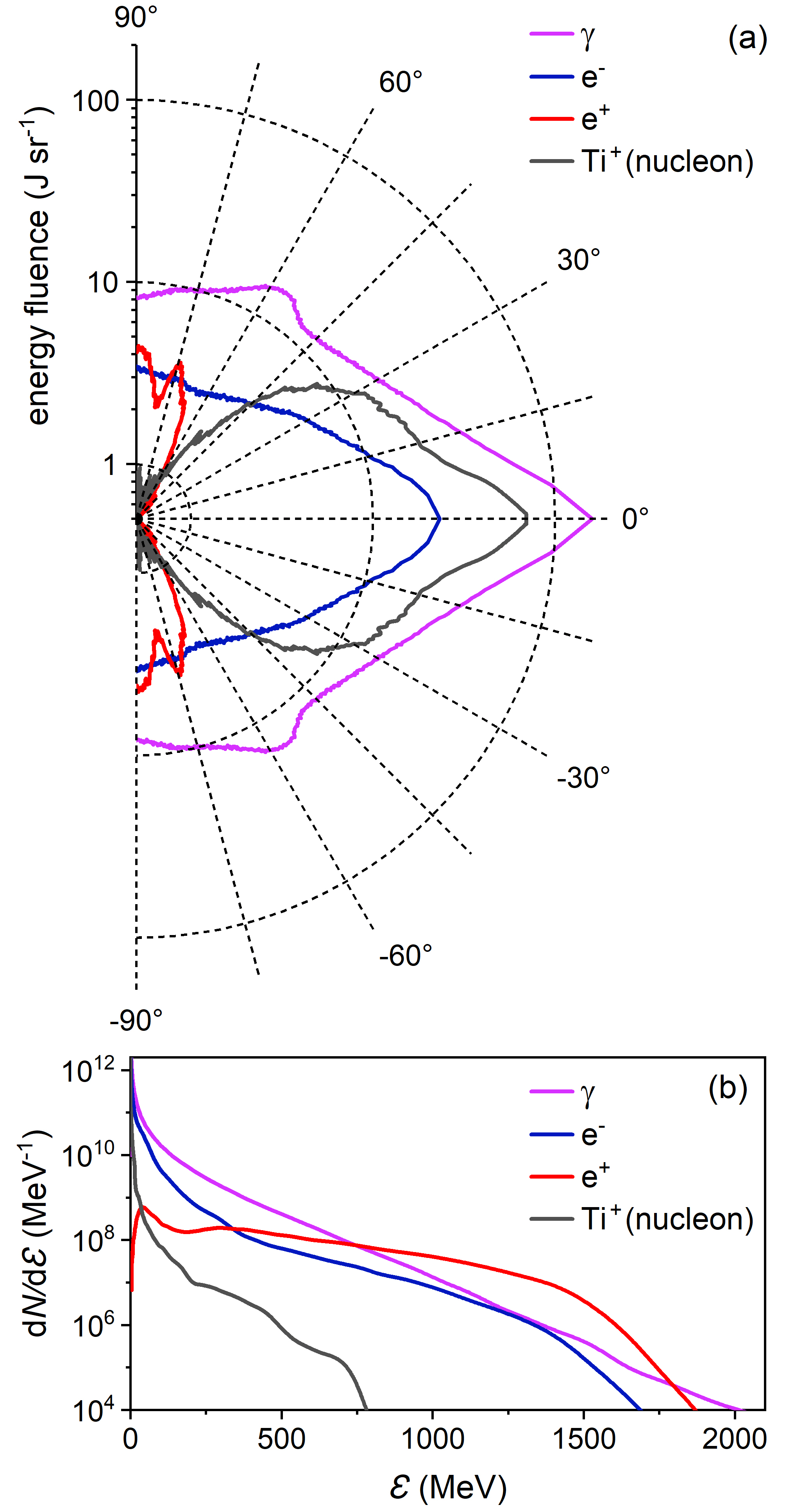}
\caption{\label{fig2} (a) The energy fluence per solid angle and (b) energy spectra of PIC simulation output particles accelerated in the forward direction and imported to the MC simulation as primary particles. \textgamma-photons, electrons, positrons, and fully ionized titanium ions (energy and particle number are per nucleon) are denoted by the violet, blue, red, and green the color, respectively. The distributions of all the particle species exhibit a cylindrical symmetry with respect to the laser propagation direction owing to the radial polarization of the laser pulse.}

\end{figure}

 As we see from Fig.~\ref{fig2}, the PIC simulation of the ultra-intense laser-matter interaction under the aforementioned conditions results in the generation of high-energy \textgamma-photons, electrons, positrons, and titanium ions, which are assumed to be fully ionized. The distributions of all the particle species exhibit a cylindrical symmetry with respect to the laser propagation direction owing to the radial polarization of the laser pulse. Approximately 57\% and 26\% of the laser energy was transferred to the kinetic energy of the particles moving in the forward direction and in the backward direction (i.e., with a negative momentum component in the laser propagation direction), respectively. For MC simulations we employ only the particles moving in the forward direction consisting of \textgamma-photons, electrons, positrons, and titanium ions; the total numbers of these particle species are approximately $3.2 \times 10^{13}$, $4.4 \times 10^{13}$, $4.5 \times 10^{11}$, $1.3 \times 10^{12}$, respectively, and the total kinetic energies of these particle species are approximately 93, 27, 10, and 30 Joules, respectively. It is worth mentioning the temperatures of the particles with the energies $>500$ MeV: $\sim140$ MeV for the \textgamma-photons, $\sim160$ MeV for electrons, and $\sim420$ MeV for positrons. Moreover, all the particles apart from positrons are strongly collimated along with the laser propagation axis, see Fig.~\ref{fig2}(a).

\subsection{\label{sec:level2b}Monte Carlo Simulation}

\par MC simulations are performed using the FLUKA code \cite{AhdidaC,2015_BattistoniG} and FLAIR graphical interface \cite{2009_VlachoudisV}. The input file of the FLUKA simulation consists of a variable number of commands, each consisting of one or more lines, also called cards. All the options and possibilities of the FLUKA code are described in the FLUKA manual \cite{2005_FerrariA}. A user-written source routine is used to import all the PIC output particles (species, position, momentum, and weight) as MC primary particles. \textgamma-photons, electrons, positrons, and ions are simulated using $\sim1.4\times10^{8}$, $\sim1.9\times10^{8}$, $\sim1.9\times10^{6}$, and $\sim2.4\times10^{6}$ primary particles, respectively. FLUKA provides simulations results per primary particle, then the results are normalized to the real number of primaries given in Subsection~\ref{sec:level2a}, corresponding to a single laser pulse. The cylindrical lead target with a diameter of 10 cm and thickness varying from 1 mm to 10 cm is positioned at a distance of 1 mm from the focal spot coordinates, as shown in Fig.~\ref{fig1}. The given geometrical setup with a large acceptance angle of $\sim 89^{\circ}$ is designed to irradiate the lead target with almost all PIC output particles moving in the forward direction.

\par The DEFAULTS card is set to PRECISIOn with the particle transport threshold set at 100 keV (particles below the threshold are not transported and their energy is deposited on the spot), except neutrons ($10^{-5} \text{ eV}$). Further models of nuclear interactions explicitly added on top of the defaults include the activation of gamma and electron interactions with nuclei, the Relativistic Quantum Molecular Dynamics (RQMD) \cite{2004_AndersenV} and a Dual Parton Model to simulate heavy ion interactions (DPMJET) \cite{2001_RoeslerS}, the evaporation and coalescence of nuclei during nuclear interactions, and the treatment of deuterons as two separate nucleons.

\par To detect the energy and angular distributions of particles passing a chosen plane, we use the USRBDX card. Employing the USRYIELD card, we obtain the temporal profile of the particles passing the given plane. Furthermore, we apply the RESNUCLE card to compute the generated residual nuclei fully de-excited down to the ground or isomeric state. For definition and description of cards used here, see the FLUKA manual in \cite{2005_FerrariA}.


\section{\label{sec:level3}Results and Discussion}

In this section, we show and discuss the results of the MC simulations of laser-driven high energy particles interacting with the lead target. The results are divided into the following subsections: the target thickness dependence of the generated nuclides and emitted particles, the energy and angular distribution of the emitted positrons, the temporal profile of collimated pulse of ultra-relativistic positrons, and the distribution of generated residual nuclides. Laser-driven particles accelerated in the forward direction resulting from the PIC simulation are referred to as primary particles in the whole section.

\subsection{\label{sec:level3a}Target thickness dependence of generated nuclides and emitted particles}

\par
\begin{figure*}[ht]
\includegraphics[width=0.94\textwidth]{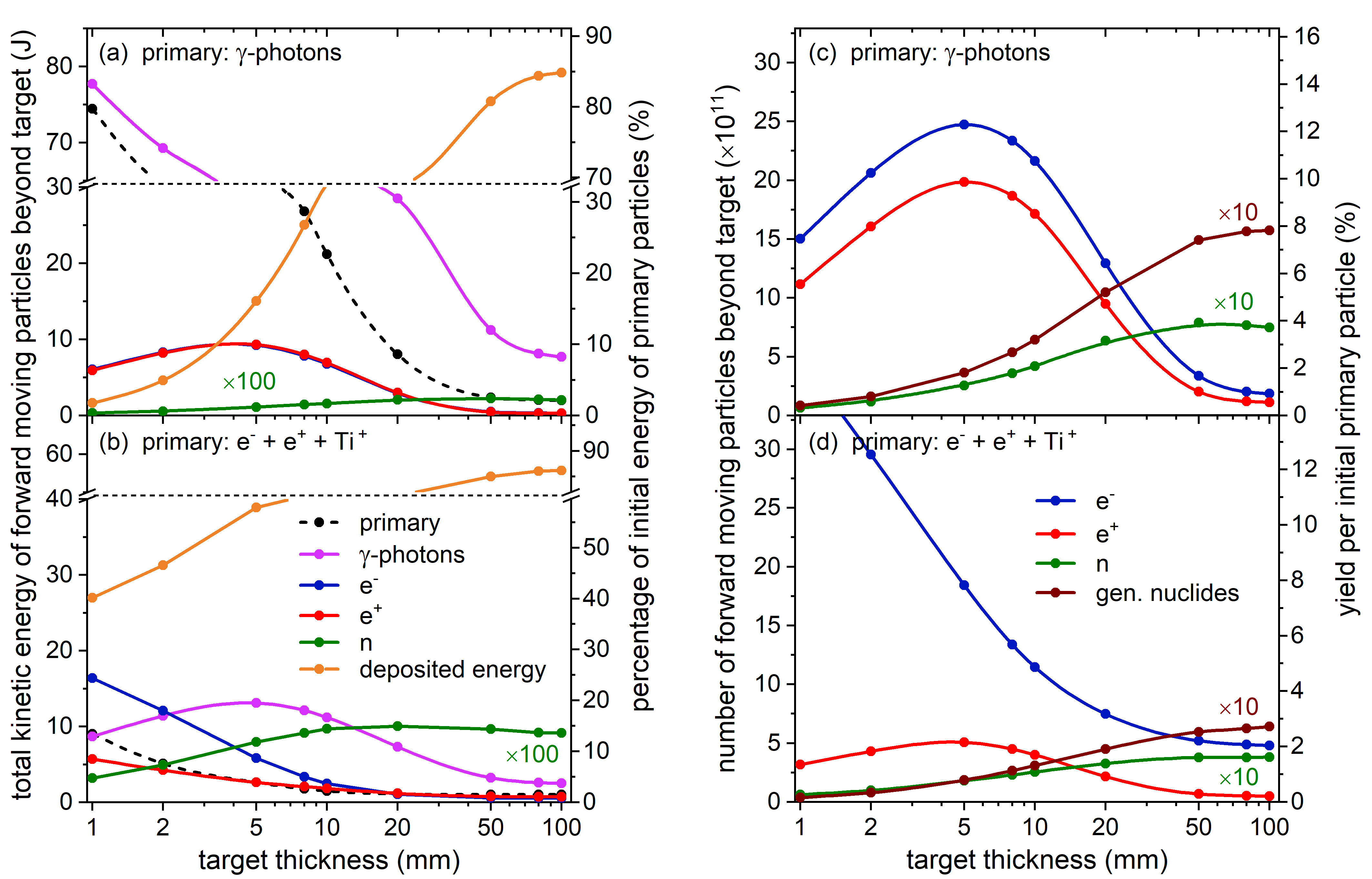}
\caption{\label{fig3}The total energy and the number of particles moving in the forward direction beyond the target as function of the thickness of the lead target. Irradiation by \textgamma-photon primary particles only are plotted in panels (a) and (c), respectively, and irradiation by the remaining primary particles (i.e., electrons, positrons, and $\text{Ti}^{+}$ ions) are plotted in panels (b) and (d), respectively. Panels (a) and (c) show also energy deposited to the target; panels (c) and (d) show the total number of residual nuclides in the target. The percentage of the initial energy of primary particles and the yield per initial primary particle are displayed on the right-hand side vertical axes. The primary particles irradiating the target are denoted with the black dashed lines; \textgamma-photons, electrons, positrons, neutrons, energy deposited to the target, and generated nuclides are denoted with the violet, blue, red, green, orange, and the vine solid lines, respectively. Note the multiplying factors for the neutrons and generated nuclides.}
\end{figure*}

 The kinetic energy of the laser-accelerated particles interacting with the solid matter can be transformed to electron-positron pair production and transferred to other particles. The particles with kinetic energy lower than the transport threshold are assumed to deposit their energy at the position of their production. As we see from panels (a) and (b) in Fig.~\ref{fig3}, the deposited energy from all primary particle species saturates at the thickness of the lead target of 10 cm.

\par The total number of electrons and positrons emitted in the forward direction from the target irradiated with \textgamma-photons plotted in Fig.~\ref{fig3} (c) implies that the majority of the generated electrons originate from the pair production. The ionization process generates predominantly low energy electrons, since the cross section increases with the decreasing \textgamma-photon energy. For \textgamma-photons of the energies below 1 MeV, the ionization becomes the dominant interaction. Although many electrons are excited to the state of a free electron by ionization, their kinetic energy is usually not high enough to escape the target. As a consequence, the ratio between electrons and positrons total kinetic energies is close to unity for all the simulated thicknesses, see Fig.~\ref{fig3} (a). The target thickness of 5 mm is optimal for the conversion of primary \textgamma-photons to positrons escaping the target in the forward direction, where both the conversion efficiency of total kinetic energy and the yield per primary \textgamma-particle reach $10\%$. Taking into account the charged primary particle species as well, plotted in panels (b) and (d) in Fig.~\ref{fig3}, the values of the total number and the total kinetic energy are higher for electrons than for positrons as a result of a considerable high number of primary electrons penetrating the target. However, these values decrease with the increasing target thickness because of its shielding effect.

\par The primary \textgamma-photons are most efficient in the generation of neutrons escaping the target and residual nuclides staying in the target for all target thicknesses, see the data plotted in panels (c) and (d) in Fig.~\ref{fig3}. The chart of generated nuclides is given in Subsection~\ref{sec:level3d}. While neutrons and residual nuclides can be produced in ions interactions, their vast majority are produced via photonuclear interactions, either directly by primary \textgamma-photons or indirectly by secondary \textgamma-photons. The secondary \textgamma-photons play an important role for photonuclear interactions since the cross section peak for the total GDR photoabsorption is narrow with FWHM of $\sim4 \text{ MeV}$ and peaks at the \textgamma-particle energy of $\sim13 \text{ MeV}$ for lead. Although the integral cross section for the total GDR photoabsorption is proportional to $NZ/(N+Z)$ \cite{1970_HaywardE}, where N is the number of neutrons, the narrow energy range stays in the same order of magnitude for all elements. Since the Bremsstrahlung from high-energy electrons and positrons is the major process of secondary \textgamma-photons generation, the primary \textgamma-photons require a relatively high target thickness of 1 cm for the maximum number of emitted secondary \textgamma-photons, while the primary charged particles reach the maximum number of emitted secondary \textgamma-photons already at 5 mm.

\par The primary \textgamma-photons irradiating the target with the thickness of 5 cm optimal for the neutron emission produced 71\% of all the neutrons emitted from the target and 19\% of the total kinetic energy of all the neutrons emitted from the target, see Fig.~\ref{fig3}. The latter is due to the aforementioned narrow cross section of the GDR process at relatively low \textgamma-photon energies. The remaining neutrons, especially the high energy ones, are generated mostly by titanium ions. High-flux neutron pulse sources find potential applications in nuclear waste management \cite{2001_RevolJP, 2008_LensaW} and material science as neutron resonance spectroscopy \cite{2010_HigginsonDP,2020_ZimmerM} and neutron diffraction \cite{2020_VogelS}. Photonuclear interaction of laser-driven \textgamma-rays for neutron source generation is simulated using PIC and MC in \cite{2015_NakamuraT}.

\subsection{\label{sec:level3b}Energy and angular distribution of emitted positrons}

\par
\begin{figure*}[ht]
\includegraphics[width=0.94\textwidth]{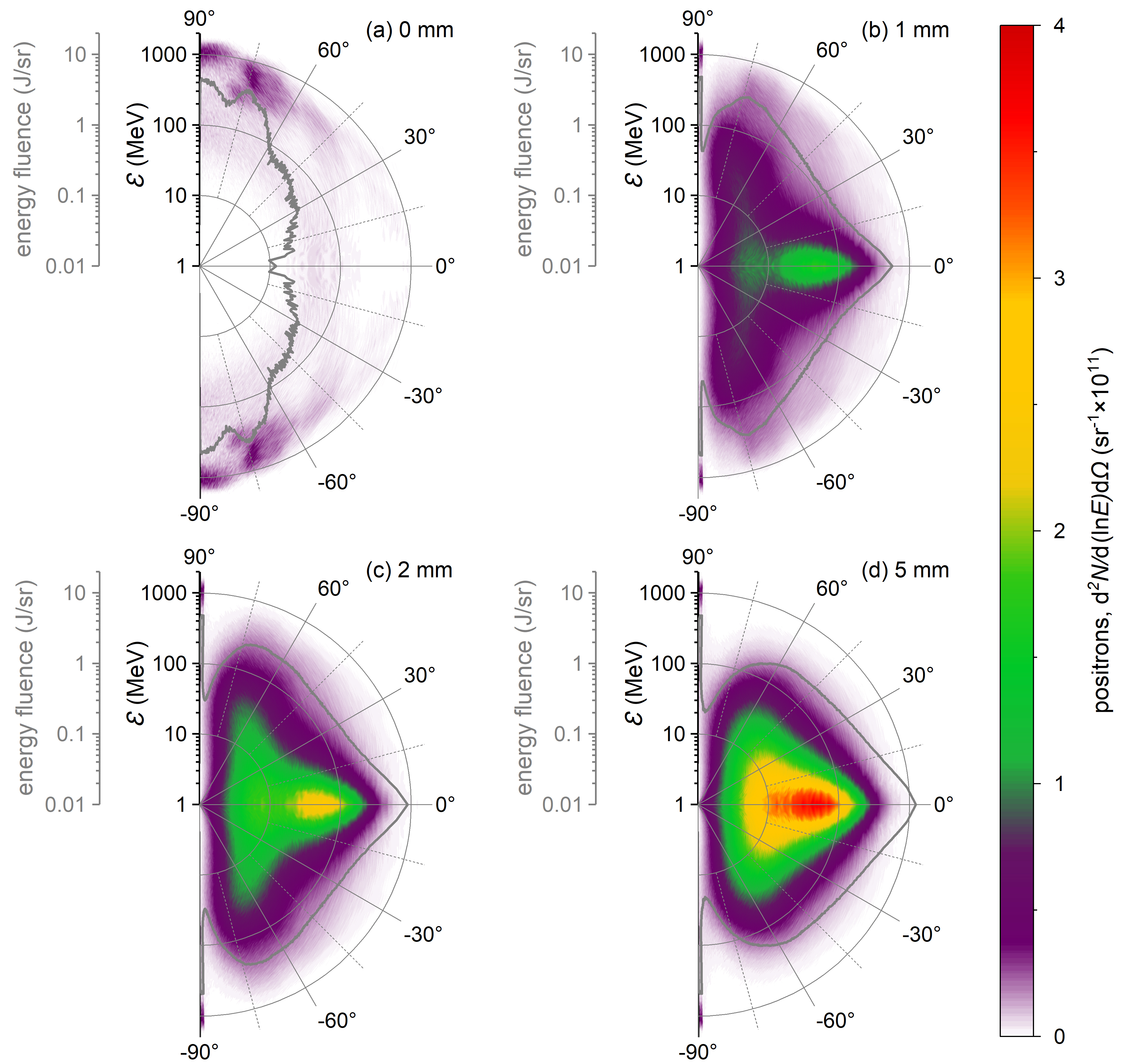}
\caption{\label{fig4}The energy and angular distributions of positrons moving in the forward direction for the lead target thickness of (a) 0 mm, (b) 1 mm, (c) 2 mm, and (d) 5mm are denoted by color mapping. The energy fluences of positrons moving in the forward direction per solid angle are overplotted with gray lines. The cylindrical symmetry of the primary particles with respect to the laser propagation direction is transferred to all emitted positrons from the target.}
\end{figure*}

 Fig.~\ref{fig4} panel (a) shows the energy and angular distribution of positrons moving in the forward direction from the PIC output. These high energy primary positrons are mostly deviated by $60^{\circ}-90^{\circ}$ from the laser propagation direction with a total number of particles of $1.6\times10^{11}$ and the total kinetic energy of 10.3 J. The major fraction of primary positrons is already shielded by a target thickness of 1 mm. On the other hand, irradiation of the target by any primary particle species can result in the generation of positrons escaping the target in the forward direction. 

\par The total kinetic energy of positrons emitted in the forward direction reaches the highest value of 12.5 J for the target thickness of 2 mm; however, a significant fraction of $\sim29\%$ of this energy comes from non-collimated primary positrons. The target thickness of 5 mm leads to the maximum number of positrons emitted in the forward direction with the value of $2.49\times10^{12}$, which is by a factor of $\sim16$ higher than the number of primary positrons. The electromagnetic cascade showers do not occur since this target thickness is comparable to the radiation length of lead of $\sim 5 \text{ mm}$. Moreover, the kinetic energy fluence per solid angle along the laser propagation axis is the highest for a target thickness of 5 mm as well; the total kinetic energy of 12 J is still higher than the total kinetic energy of primary positrons by a factor of $\sim1.2$. The spectrum of all positrons emitted from the 5 mm thick target in the forward direction exhibits two temperatures. The population with an energy below 500 MeV, mostly secondaries produced by primary \textgamma-photons and electrons, has a temperature of $\sim80 \text{ MeV}$. The population with an energy above 500 MeV mostly formed by primary positrons has a temperature of $\sim410 \text{ MeV}$.

\par It is shown in Fig.~\ref{fig3} that the major fraction of positrons emitted from the target in the forward direction and the major fraction of total kinetic energy of positrons emitted from the target in the forward direction are caused by target irradiation with primary \textgamma-photons for all simulated target thicknesses. The charged particles are less efficient since they have to first produce a Bremsstrahlung \textgamma-photon to generate electron-positron pair. In particular, for the target thickness of 5 mm optimal for positron emission, the primary \textgamma-photons, electrons, and positrons are responsible for $\sim80\%$, $\sim10\%$, and $\sim10\%$ of positrons escaping the target, respectively; titanium ions produce an insignificant number of positrons. According to the doubly differential Bethe-Heitler cross section for the electron-positron pair production \cite{2014_KoehnCh}, the most probable scattering angle between incident \textgamma-photon and produced positron decreases with the increasing \textgamma-photon energy and the positron fraction of total pair kinetic energy. For \textgamma-photon energies $< 10\text{ MeV}$, the most probable scattering angles of produced positrons are $>5^{\circ}$. Nevertheless, the peak of the angular distribution of the scattering probability is asymmetric with a slower decrease towards the smaller angles than the bigger angles. 

\par Owing to the aforementioned properties of electron-positron pair production cross section and strongly collimated energy fluence of primary \textgamma-photons, see Fig.~\ref{fig2}, the highest fluences per solid angle of positrons escaping the target and of their total kinetic energy are computed in the laser propagation direction for all target thicknesses plotted in panels (b), (c), and (d) in Fig.~\ref{fig4}. The highest population of resulting spectra along with the laser propagation are in the energy range of $10 - 100\text{ MeV}$.

\subsection{\label{sec:level3c}Temporal profile of collimated pulse of ultra-relativistic positrons}

\par
\begin{figure}[h]
\includegraphics[width=0.47\textwidth]{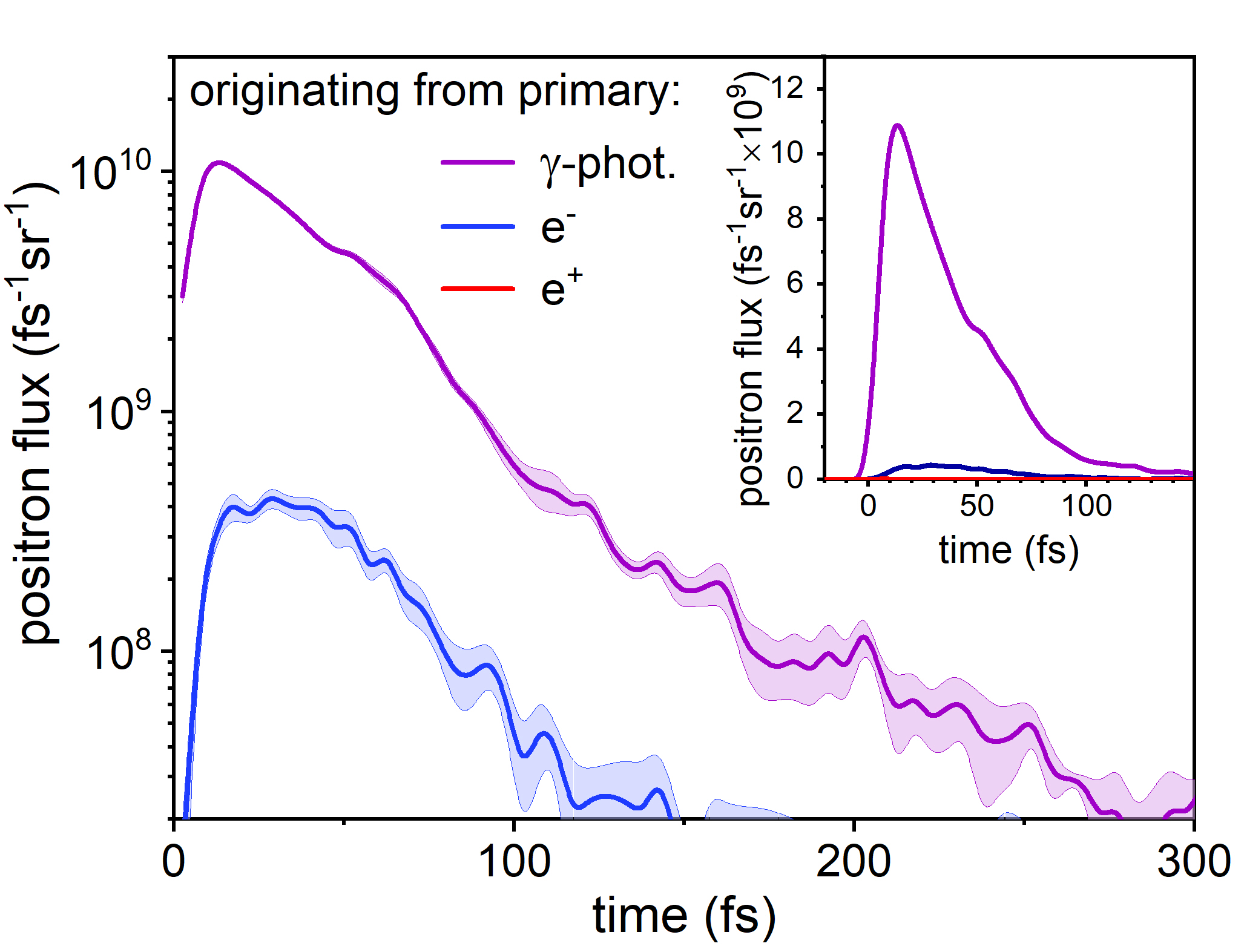}
\caption{\label{fig5}The flux of positrons per solid angle escaping from a 5 mm thick target in the forward direction and entering the disc detector with a diameter of 1 mm aligned concentrically and in parallel to the target at a distance of 1 mm. The minimum energy threshold is 10 MeV and the maximum threshold of divergence angle from the target normal is 50 mrad. The data for positrons originating from primary \textgamma-photons, electrons, and positrons are denoted with violet, blue, and red color, respectively; the bands represent errors of estimated values. The inset figure displays the results with a linear scale. The zero point of the time axes is placed at the first detection of positrons. Note the positron data are below the logarithmic range of the vertical axis in the main panel and overlap the horizontal axis in the inset.}
\end{figure}

 Fig.~\ref{fig5} shows the pulse profile of positrons with the energy $>10 \text{ MeV}$ escaping a 5 mm thick target and entering the disc detector with a maximum divergence of 50 mrad from the target normal; the detector with a diameter of 1 mm is aligned concentrically and in parallel to the target at a distance of 1 mm. These thresholds and target-to-detector geometry are designed for the detection of ultra-relativistic positrons collimated with the laser propagation axis that can be further focused or moderated or both for the requirements of a particular application. The detected positron pulse exhibits a short duration of 35 fs (FWHM) and predominantly originates from primary \textgamma-photons, as expected from the previous discussion. 
The positrons decay exponentially with the time constant of $\sim43 \text{ fs}$ and $\sim34 \text{ fs}$ for positrons originating from primary \textgamma-photons and electrons, respectively. Calculated brightness of positron beam is $\sim1.1 \kern0.2em \mathrm{A/mm^{2}/mrad^{2}}$.

\par The data for electrons and positrons in Fig.~\ref{fig3} implies that the ratio between high energy electrons and positrons emitted from the target converges to unity with increasing target thickness owing to the higher shielding effect for primary electrons and positrons. For all simulated target thicknesses, see Subsection~\ref{sec:level3a}, the dominant process of electron generation is electron-positron pair production which is a symmetric interaction in electron and positron angular and energy distributions. By neglecting the scattering effect, the most probable mean opening angle in radians is $\theta = 1.6/k_{\text{MeV}}$, where $k_{\text{MeV}}$ is the energy of pair generating \textgamma-photon in MeV \cite{1963_OlsenH}. As an example, the average electron and positron both with energy of 10 MeV are emitted with an opening angle of $\sim80 \text{ mrad}$, i.e. $\sim4.6 \text{ deg}$. Owing to the strong collimation of primary high energy \textgamma-photons \cite{2021_HadjisolomouP}, the emission of high-energy electrons and positrons with low mean opening angles is predominantly directed along the laser propagation axis. The pulse of electrons and positrons emitted from the target with the presented properties can be characterized as high-flux relativistic pair jets \cite{2015_ChenH} and used for the electron-positron pair plasmas studies, where apart from other criteria the high and equal densities of electrons and positrons in the volume of their co-existence are required to achieve quasi-neutrality and collective behavior \cite{2010_ChenH, 2011_ChenH,2015_SarriG}. Studies on high-flux relativistic pair jets and electron-positron pair plasmas open new areas of research, particularly in laboratory astrophysics.

\subsection{\label{sec:level3d}Distribution of generated residual nuclides}

\par
\begin{figure}[h]
\includegraphics[width=0.47\textwidth]{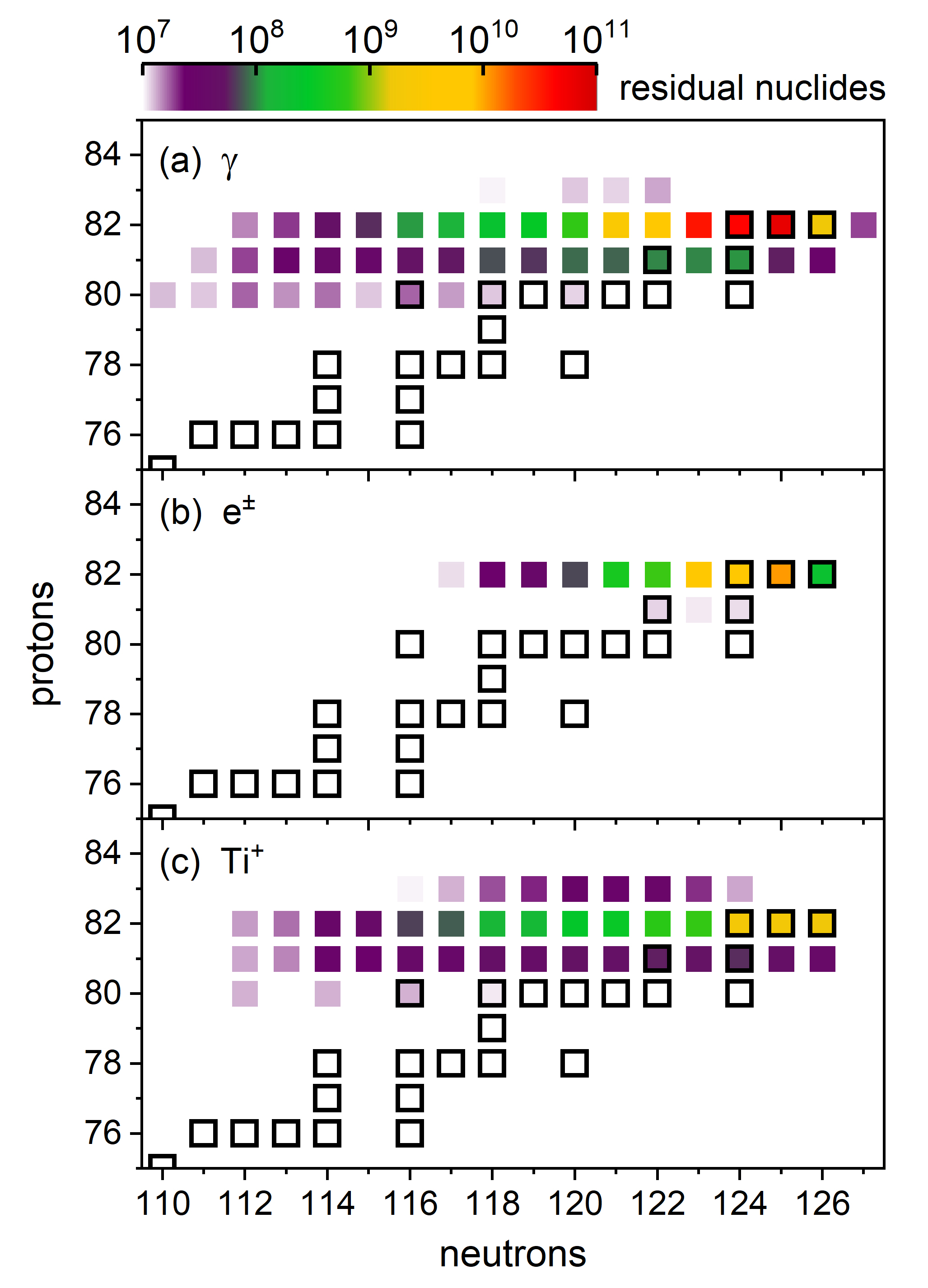}
\caption{\label{fig6}Chart of residual high-Z nuclides generated in the lead target with a thickness of 10 cm irradiated by primary a) \textgamma-photons, b) electrons and positrons, and c) titanium ions from the single laser pulse. Stable nuclides are indicated with squares with black edges.}
\end{figure}

 The results of the MC simulation show that all species of primary particles irradiating the lead target produce not only stable but also radioactive residual nuclides. Fig.~\ref{fig6} displays the numbers of residual high-Z nuclides created in the target with the highest simulated thickness of 10 cm irradiated by primary \textgamma-photons, electrons/positrons, and titanium ions, individually. The natural lead used as a target is comprised of three stable isotopes denoted by squares with black edges in the row $Z = 82$ with a standard atomic weight of $A = 207.19$. The majority of residual nuclides are produced through photonuclear processes, predominantly through GDR, either directly by primary \textgamma-photons or indirectly by secondary Bremsstrahlung \textgamma-photons from fast electrons and positrons, see Subsection~\ref{sec:level3a}. The remaining minor processes generating residual nuclides are various nucleus-nucleus interactions with heavy ions and electonuclear interactions.

\par Among all generated nuclides in the lead target, the most abundant ones are lead isotopes, protons, and alpha particles. One of the most numerous lead radioactive isotopes generated is $^{203}{\rm Pb}$ where $\sim4\times10^{9}$ nuclides are produced by all primary particles from a single laser pulse. The direct decay of $^{203}{\rm Pb}$ to stable $^{203}_{81}{\rm Tl}$ through electron capture with half-life of $\sim52$ hours, emitted \textgamma-photon energy of $\sim279.2 \text{ keV}$, and the absence of hadron emission makes it particularly suitable for medical imaging \cite{2014_AzzamA}. High number of thallium isotopes ($Z = 81$) is also generated in the target with a total count of $\sim10^{9}$ nuclei per single laser pulse. Among them, the most useful radioisotope being historically used extensively for nuclear medicine is $^{201}{\rm Tl}$ where $\sim1.5\times10^{8}$ nuclides are produced. The long half-life of $\sim73$ hours and the direct decay to stable $^{201}_{80}{\rm Hg}$ by electron capture provide convenient shelf storage and successful imaging over a period of hours \cite{1999_TadamuraE}. Other interesting radioactive isotopes of thallium produced for medical applications are $^{206}{\rm Tl}$, $^{200}{\rm Tl}$, $^{199}{\rm Tl}$, $^{198}{\rm Tl}$, and $^{197}{\rm Tl}$ with half-lives of 4.2 min, 26.1 hours, 7.4 h, 5.3 hours, and 2.8 hours, respectively.


\section{\label{sec:level4}Conclusions}

\par We have examined the interaction of the \textgamma-photon flash and high energy charged particles generated by an ultra-intense laser in the $\lambda^3$ regime with a high-Z target by using a 3D PIC (EPOCH) simulation coupled with MC (FLUKA) simulations. We have found that the interaction of laser-driven high energy particles with the cylindrical lead target can produce an ultra-short pulse of ultra-relativistic positrons emitted from the target collimated to the laser propagation axis and a broad source of radioactive nuclides. The highest population of resulting spectra along with the laser propagation axis are in the energy range of $10 - 100\text{ MeV}$. The computed pulse of ultra-relativistic positrons with energies $>10 \text{ MeV}$ and the divergence angle from the target normal $< 50 \text{ mrad}$ originate predominantly from primary \textgamma-photons and exhibit a very short duration of 35 fs (FWHM). 

\par Such an ultra-short and ultra-relativistic positron beam can be further modified and utilized as a probe for high-resolution and volumetric scanning of materials using positron annihilation lifetime spectroscopy \cite{2016_XuT,2021_AudetTL,2021_SarriG}. The ultra-short duration of the pulsed positron sources is necessary for fine temporal resolution in annihilation lifetime measurement since the pulse duration should be significantly smaller than a timescale of interest. The energy range of tens and hundreds of MeV can extend this technique to the study of millimeter-centimeter thick materials. Further modified properties of the collimated source of high energy positrons can find a potential application as the injector in laser-based accelerators of charged particles. Laser-driven positron beams can potentially become injectors of reduced cost and size for future high energy accelerators and lepton colliders \cite{2019_AlejoA}. The ultra-relativistic positrons are accompanied by electrons in the same direction as the opening angles for ultra-relativistic pairs are very low. The pulse of electrons and positrons emitted from the target with presented characteristics can be used for the electron-positron pair plasmas studies \cite{2010_ChenH, 2011_ChenH,2015_SarriG} opening new areas of research, particularly in laboratory astrophysics research. The properties such as high density of pairs due to the short pulse duration, high energy of pairs reaching hundreds of MeV, and low opening angle emission, characterize the electron-positron pulses, referred to as high-flux relativistic pair jets \cite{2015_ChenH}, well suited for the study of scaled high energy astrophysical phenomena including relativistic shocks formation in laboratory experiments.

\par We have also investigated the target thickness dependence of the number and total kinetic energy of various particles emitted from the lead target irradiated by laser-generated \textgamma-photons and charged particles separately. Such high-intensity neutron pulse sources can find potential applications in nuclear waste management \cite{2001_RevolJP, 2008_LensaW} and material science as neutron resonance spectroscopy \cite{2010_HigginsonDP,2020_ZimmerM} for probing the interior ion temperature of opaque materials and neutron diffraction \cite{2020_VogelS} for analysis of the atomic and magnetic structure of materials. The short pulse duration and the ability to dope only a given section of the material enable high temporal and spatial resolution of the analysis. Owing to the short pulse duration the continuous spectrum can be used with the time-of-flight detection method.

\par The majority of residual nuclides are produced through photonuclear processes dominantly by the giant dipole resonance. Activation of residual nuclides can be used for the study of nuclear physics and astrophysics, and for direct applications in nuclear medicine. For nuclear medicine, the important radioactive isotopes produced are $^{203}{\rm Pb}$ and $^{201}{\rm Tl}$ with a total count of $\sim4\times10^{9}$ and $\sim1.5\times10^{8}$ nuclei produced in a 10 cm thick lead target by all primary particles per single laser pulse. These nuclides undergo direct decay to stable nuclides through electron capture with a half-life of tens of hours. They emit \textgamma-photons of suitable energies for medical imaging.


\begin{acknowledgments}
This work is supported by the projects High Field Initiative (CZ.02.1.01/0.0/0.0/15\_003/0000449) from the European Regional Development Fund and “e-INFRA CZ” (ID:90140) from the Ministry of Education, Youth and Sports of the Czech Republic. The EPOCH code is in part funded by the UK EPSRC grants EP/G054950/1, EP/G056803/1, EP/G055165/1 and EP/M022463/1.
 \end{acknowledgments}



\bibliography{KolenatyD_manuscript}

\providecommand{\noopsort}[1]{}\providecommand{\singleletter}[1]{#1}%
\begin{thebibliography}{56}%
\makeatletter
\providecommand \@ifxundefined [1]{%
 \@ifx{#1\undefined}
}%
\providecommand \@ifnum [1]{%
 \ifnum #1\expandafter \@firstoftwo
 \else \expandafter \@secondoftwo
 \fi
}%
\providecommand \@ifx [1]{%
 \ifx #1\expandafter \@firstoftwo
 \else \expandafter \@secondoftwo
 \fi
}%
\providecommand \natexlab [1]{#1}%
\providecommand \enquote  [1]{``#1''}%
\providecommand \bibnamefont  [1]{#1}%
\providecommand \bibfnamefont [1]{#1}%
\providecommand \citenamefont [1]{#1}%
\providecommand \href@noop [0]{\@secondoftwo}%
\providecommand \href [0]{\begingroup \@sanitize@url \@href}%
\providecommand \@href[1]{\@@startlink{#1}\@@href}%
\providecommand \@@href[1]{\endgroup#1\@@endlink}%
\providecommand \@sanitize@url [0]{\catcode `\\12\catcode `\$12\catcode
  `\&12\catcode `\#12\catcode `\^12\catcode `\_12\catcode `\%12\relax}%
\providecommand \@@startlink[1]{}%
\providecommand \@@endlink[0]{}%
\providecommand \url  [0]{\begingroup\@sanitize@url \@url }%
\providecommand \@url [1]{\endgroup\@href {#1}{\urlprefix }}%
\providecommand \urlprefix  [0]{URL }%
\providecommand \Eprint [0]{\href }%
\providecommand \doibase [0]{https://doi.org/}%
\providecommand \selectlanguage [0]{\@gobble}%
\providecommand \bibinfo  [0]{\@secondoftwo}%
\providecommand \bibfield  [0]{\@secondoftwo}%
\providecommand \translation [1]{[#1]}%
\providecommand \BibitemOpen [0]{}%
\providecommand \bibitemStop [0]{}%
\providecommand \bibitemNoStop [0]{.\EOS\space}%
\providecommand \EOS [0]{\spacefactor3000\relax}%
\providecommand \BibitemShut  [1]{\csname bibitem#1\endcsname}%
\let\auto@bib@innerbib\@empty
\bibitem [{\citenamefont {Strickland}\ and\ \citenamefont
  {Mourou}(1985)}]{1985_StricklandD}%
  \BibitemOpen
  \bibfield  {author} {\bibinfo {author} {\bibfnamefont {D.}~\bibnamefont
  {Strickland}}\ and\ \bibinfo {author} {\bibfnamefont {G.}~\bibnamefont
  {Mourou}},\ }\bibfield  {title} {\bibinfo {title} {{Compression of amplified
  chirped optical pulses}},\ }\href
  {https://doi.org/10.1016/0030-4018(85)90151-8} {\bibfield  {journal}
  {\bibinfo  {journal} {Opt. Commun.}\ }\textbf {\bibinfo {volume} {55}},\
  \bibinfo {pages} {447} (\bibinfo {year} {1985})}\BibitemShut {NoStop}%
\bibitem [{\citenamefont {Spence}\ \emph {et~al.}(1991)\citenamefont {Spence},
  \citenamefont {Kean},\ and\ \citenamefont {Sibbett}}]{1991_SpenceDE}%
  \BibitemOpen
  \bibfield  {author} {\bibinfo {author} {\bibfnamefont {D.~E.}\ \bibnamefont
  {Spence}}, \bibinfo {author} {\bibfnamefont {P.~N.}\ \bibnamefont {Kean}},\
  and\ \bibinfo {author} {\bibfnamefont {W.}~\bibnamefont {Sibbett}},\
  }\bibfield  {title} {\bibinfo {title} {{60-fsec pulse generation from a
  self-mode-locked Ti:sapphire laser}},\ }\href
  {https://doi.org/10.1364/OL.16.000042} {\bibfield  {journal} {\bibinfo
  {journal} {Opt. Lett.}\ }\textbf {\bibinfo {volume} {16}},\ \bibinfo {pages}
  {42} (\bibinfo {year} {1991})}\BibitemShut {NoStop}%
\bibitem [{\citenamefont {Danson}\ \emph {et~al.}(2019)\citenamefont {Danson},
  \citenamefont {Haefner}, \citenamefont {Bromage}, \citenamefont {Butcher},
  \citenamefont {Chanteloup}, \citenamefont {Chowdhury}, \citenamefont
  {Galvanauskas}, \citenamefont {Gizzi}, \citenamefont {Hein}, \citenamefont
  {Hillier},\ and\ \citenamefont {et~al.}}]{2019_DansonCN}%
  \BibitemOpen
  \bibfield  {author} {\bibinfo {author} {\bibfnamefont {C.~N.}\ \bibnamefont
  {Danson}}, \bibinfo {author} {\bibfnamefont {C.}~\bibnamefont {Haefner}},
  \bibinfo {author} {\bibfnamefont {J.}~\bibnamefont {Bromage}}, \bibinfo
  {author} {\bibfnamefont {T.}~\bibnamefont {Butcher}}, \bibinfo {author}
  {\bibfnamefont {J.-C.~F.}\ \bibnamefont {Chanteloup}}, \bibinfo {author}
  {\bibfnamefont {E.~A.}\ \bibnamefont {Chowdhury}}, \bibinfo {author}
  {\bibfnamefont {A.}~\bibnamefont {Galvanauskas}}, \bibinfo {author}
  {\bibfnamefont {L.~A.}\ \bibnamefont {Gizzi}}, \bibinfo {author}
  {\bibfnamefont {J.}~\bibnamefont {Hein}}, \bibinfo {author} {\bibfnamefont
  {D.~I.}\ \bibnamefont {Hillier}},\ and\ \bibinfo {author} {\bibnamefont
  {et~al.}},\ }\bibfield  {title} {\bibinfo {title} {{Petawatt and exawatt
  class lasers worldwide}},\ }\href {https://doi.org/10.1017/hpl.2019.36}
  {\bibfield  {journal} {\bibinfo  {journal} {High Power Laser Sci. Eng.}\
  }\textbf {\bibinfo {volume} {7}},\ \bibinfo {pages} {54} (\bibinfo {year}
  {2019})}\BibitemShut {NoStop}%
\bibitem [{\citenamefont {Tanaka}\ \emph {et~al.}(2020)\citenamefont {Tanaka},
  \citenamefont {Spohr}, \citenamefont {Balabanski}, \citenamefont {Balascuta},
  \citenamefont {Capponi}, \citenamefont {Cernaianu}, \citenamefont {Cuciuc},
  \citenamefont {Cucoanes}, \citenamefont {Dancus}, \citenamefont {Dhal},\ and\
  \citenamefont {et~al.}}]{2020_TanakaKA}%
  \BibitemOpen
  \bibfield  {author} {\bibinfo {author} {\bibfnamefont {K.~A.}\ \bibnamefont
  {Tanaka}}, \bibinfo {author} {\bibfnamefont {K.~M.}\ \bibnamefont {Spohr}},
  \bibinfo {author} {\bibfnamefont {D.~L.}\ \bibnamefont {Balabanski}},
  \bibinfo {author} {\bibfnamefont {S.}~\bibnamefont {Balascuta}}, \bibinfo
  {author} {\bibfnamefont {L.}~\bibnamefont {Capponi}}, \bibinfo {author}
  {\bibfnamefont {M.~O.}\ \bibnamefont {Cernaianu}}, \bibinfo {author}
  {\bibfnamefont {M.}~\bibnamefont {Cuciuc}}, \bibinfo {author} {\bibfnamefont
  {A.}~\bibnamefont {Cucoanes}}, \bibinfo {author} {\bibfnamefont
  {I.}~\bibnamefont {Dancus}}, \bibinfo {author} {\bibfnamefont
  {A.}~\bibnamefont {Dhal}},\ and\ \bibinfo {author} {\bibnamefont {et~al.}},\
  }\bibfield  {title} {\bibinfo {title} {{Current status and highlights of the
  ELI-NP research program}},\ }\href {https://doi.org/10.1063/1.5093535}
  {\bibfield  {journal} {\bibinfo  {journal} {Matter Radiat. at Extremes}\
  }\textbf {\bibinfo {volume} {5}},\ \bibinfo {pages} {024402} (\bibinfo {year}
  {2020})}\BibitemShut {NoStop}%
\bibitem [{\citenamefont {Yoon}\ \emph {et~al.}(2021)\citenamefont {Yoon},
  \citenamefont {Kim}, \citenamefont {Choi}, \citenamefont {Sung},
  \citenamefont {Lee}, \citenamefont {Lee},\ and\ \citenamefont
  {Nam}}]{2021_YoonJW}%
  \BibitemOpen
  \bibfield  {author} {\bibinfo {author} {\bibfnamefont {J.~W.}\ \bibnamefont
  {Yoon}}, \bibinfo {author} {\bibfnamefont {Y.~G.}\ \bibnamefont {Kim}},
  \bibinfo {author} {\bibfnamefont {I.~W.}\ \bibnamefont {Choi}}, \bibinfo
  {author} {\bibfnamefont {J.~H.}\ \bibnamefont {Sung}}, \bibinfo {author}
  {\bibfnamefont {H.~W.}\ \bibnamefont {Lee}}, \bibinfo {author} {\bibfnamefont
  {S.~K.}\ \bibnamefont {Lee}},\ and\ \bibinfo {author} {\bibfnamefont {C.~H.}\
  \bibnamefont {Nam}},\ }\bibfield  {title} {\bibinfo {title} {{Realization of
  laser intensity over $10^{23} \text{ W/cm}^{2}$}},\ }\href
  {https://doi.org/10.1364/OPTICA.420520} {\bibfield  {journal} {\bibinfo
  {journal} {Optica}\ }\textbf {\bibinfo {volume} {8}},\ \bibinfo {pages} {630}
  (\bibinfo {year} {2021})}\BibitemShut {NoStop}%
\bibitem [{\citenamefont {Mourou}\ \emph {et~al.}(2002)\citenamefont {Mourou},
  \citenamefont {Chang}, \citenamefont {Maksimchuk}, \citenamefont {Nees},
  \citenamefont {Bulanov}, \citenamefont {Bychenkov}, \citenamefont
  {Esirkepov}, \citenamefont {Naumova}, \citenamefont {Pegoraro},\ and\
  \citenamefont {Ruhl}}]{2002_MourouG}%
  \BibitemOpen
  \bibfield  {author} {\bibinfo {author} {\bibfnamefont {G.}~\bibnamefont
  {Mourou}}, \bibinfo {author} {\bibfnamefont {Z.}~\bibnamefont {Chang}},
  \bibinfo {author} {\bibfnamefont {A.}~\bibnamefont {Maksimchuk}}, \bibinfo
  {author} {\bibfnamefont {J.}~\bibnamefont {Nees}}, \bibinfo {author}
  {\bibfnamefont {S.~V.}\ \bibnamefont {Bulanov}}, \bibinfo {author}
  {\bibfnamefont {V.~Y.}\ \bibnamefont {Bychenkov}}, \bibinfo {author}
  {\bibfnamefont {T.~Z.}\ \bibnamefont {Esirkepov}}, \bibinfo {author}
  {\bibfnamefont {N.~M.}\ \bibnamefont {Naumova}}, \bibinfo {author}
  {\bibfnamefont {F.}~\bibnamefont {Pegoraro}},\ and\ \bibinfo {author}
  {\bibfnamefont {H.}~\bibnamefont {Ruhl}},\ }\bibfield  {title} {\bibinfo
  {title} {{On the design of experiments for the study of relativistic
  nonlinear optics in the limit of single-cycle pulse duration and
  single-wavelength spot size}},\ }\href {https://doi.org/10.1134/1.1434292}
  {\bibfield  {journal} {\bibinfo  {journal} {Plasma Phys. Rep.}\ }\textbf
  {\bibinfo {volume} {28}},\ \bibinfo {pages} {12} (\bibinfo {year}
  {2002})}\BibitemShut {NoStop}%
\bibitem [{\citenamefont {Esarey}\ \emph {et~al.}(2009)\citenamefont {Esarey},
  \citenamefont {Schroeder},\ and\ \citenamefont {Leemans}}]{2009_EsareyE}%
  \BibitemOpen
  \bibfield  {author} {\bibinfo {author} {\bibfnamefont {E.}~\bibnamefont
  {Esarey}}, \bibinfo {author} {\bibfnamefont {C.~B.}\ \bibnamefont
  {Schroeder}},\ and\ \bibinfo {author} {\bibfnamefont {W.~P.}\ \bibnamefont
  {Leemans}},\ }\bibfield  {title} {\bibinfo {title} {{Physics of laser-driven
  plasma-based electron accelerators}},\ }\href
  {https://doi.org/10.1103/RevModPhys.81.1229} {\bibfield  {journal} {\bibinfo
  {journal} {Rev. Mod. Phys.}\ }\textbf {\bibinfo {volume} {81}},\ \bibinfo
  {pages} {1229} (\bibinfo {year} {2009})}\BibitemShut {NoStop}%
\bibitem [{\citenamefont {Daido}\ \emph {et~al.}(2012)\citenamefont {Daido},
  \citenamefont {Nishiuchi},\ and\ \citenamefont {Pirozhkov}}]{2012_DaidoH}%
  \BibitemOpen
  \bibfield  {author} {\bibinfo {author} {\bibfnamefont {H.}~\bibnamefont
  {Daido}}, \bibinfo {author} {\bibfnamefont {M.}~\bibnamefont {Nishiuchi}},\
  and\ \bibinfo {author} {\bibfnamefont {A.~S.}\ \bibnamefont {Pirozhkov}},\
  }\bibfield  {title} {\bibinfo {title} {{Review of laser-driven ion sources
  and their applications}},\ }\href
  {https://doi.org/10.1088/0034-4885/75/5/056401} {\bibfield  {journal}
  {\bibinfo  {journal} {Rep. Prog. Phys.}\ }\textbf {\bibinfo {volume} {75}},\
  \bibinfo {pages} {056401} (\bibinfo {year} {2012})}\BibitemShut {NoStop}%
\bibitem [{\citenamefont {Mourou}\ \emph {et~al.}(2006)\citenamefont {Mourou},
  \citenamefont {Tajima},\ and\ \citenamefont {Bulanov}}]{2006_MourouG}%
  \BibitemOpen
  \bibfield  {author} {\bibinfo {author} {\bibfnamefont {G.~A.}\ \bibnamefont
  {Mourou}}, \bibinfo {author} {\bibfnamefont {T.}~\bibnamefont {Tajima}},\
  and\ \bibinfo {author} {\bibfnamefont {S.~V.}\ \bibnamefont {Bulanov}},\
  }\bibfield  {title} {\bibinfo {title} {{Optics in the relativistic regime}},\
  }\href {https://doi.org/10.1103/RevModPhys.78.309} {\bibfield  {journal}
  {\bibinfo  {journal} {Rev. Mod. Phys.}\ }\textbf {\bibinfo {volume} {78}},\
  \bibinfo {pages} {309} (\bibinfo {year} {2006})}\BibitemShut {NoStop}%
\bibitem [{\citenamefont {Nakamura}\ \emph {et~al.}(2012)\citenamefont
  {Nakamura}, \citenamefont {Koga}, \citenamefont {Esirkepov}, \citenamefont
  {Kando}, \citenamefont {Korn},\ and\ \citenamefont
  {Bulanov}}]{2012_NakamuraT}%
  \BibitemOpen
  \bibfield  {author} {\bibinfo {author} {\bibfnamefont {T.}~\bibnamefont
  {Nakamura}}, \bibinfo {author} {\bibfnamefont {J.~K.}\ \bibnamefont {Koga}},
  \bibinfo {author} {\bibfnamefont {T.~Z.}\ \bibnamefont {Esirkepov}}, \bibinfo
  {author} {\bibfnamefont {M.}~\bibnamefont {Kando}}, \bibinfo {author}
  {\bibfnamefont {G.}~\bibnamefont {Korn}},\ and\ \bibinfo {author}
  {\bibfnamefont {S.~V.}\ \bibnamefont {Bulanov}},\ }\bibfield  {title}
  {\bibinfo {title} {{High-power $\ensuremath{\gamma}$-ray flash generation in
  ultraintense laser-plasma interactions}},\ }\href
  {https://doi.org/10.1103/PhysRevLett.108.195001} {\bibfield  {journal}
  {\bibinfo  {journal} {Phys. Rev. Lett.}\ }\textbf {\bibinfo {volume} {108}},\
  \bibinfo {pages} {195001} (\bibinfo {year} {2012})}\BibitemShut {NoStop}%
\bibitem [{\citenamefont {Ridgers}\ \emph {et~al.}(2012)\citenamefont
  {Ridgers}, \citenamefont {Brady}, \citenamefont {Duclous}, \citenamefont
  {Kirk}, \citenamefont {Bennett}, \citenamefont {Arber}, \citenamefont
  {Robinson},\ and\ \citenamefont {Bell}}]{2012_RidgersCP}%
  \BibitemOpen
  \bibfield  {author} {\bibinfo {author} {\bibfnamefont {C.~P.}\ \bibnamefont
  {Ridgers}}, \bibinfo {author} {\bibfnamefont {C.~S.}\ \bibnamefont {Brady}},
  \bibinfo {author} {\bibfnamefont {R.}~\bibnamefont {Duclous}}, \bibinfo
  {author} {\bibfnamefont {J.~G.}\ \bibnamefont {Kirk}}, \bibinfo {author}
  {\bibfnamefont {K.}~\bibnamefont {Bennett}}, \bibinfo {author} {\bibfnamefont
  {T.~D.}\ \bibnamefont {Arber}}, \bibinfo {author} {\bibfnamefont {A.~P.~L.}\
  \bibnamefont {Robinson}},\ and\ \bibinfo {author} {\bibfnamefont {A.~R.}\
  \bibnamefont {Bell}},\ }\bibfield  {title} {\bibinfo {title} {{Dense
  electron-positron plasmas and ultraintense $\ensuremath{\gamma}$ rays from
  laser-irradiated solids}},\ }\href
  {https://doi.org/10.1103/PhysRevLett.108.165006} {\bibfield  {journal}
  {\bibinfo  {journal} {Phys. Rev. Lett.}\ }\textbf {\bibinfo {volume} {108}},\
  \bibinfo {pages} {165006} (\bibinfo {year} {2012})}\BibitemShut {NoStop}%
\bibitem [{\citenamefont {Lezhnin}\ \emph {et~al.}(2018)\citenamefont
  {Lezhnin}, \citenamefont {Sasorov}, \citenamefont {Korn},\ and\ \citenamefont
  {Bulanov}}]{2018_LezhninKV}%
  \BibitemOpen
  \bibfield  {author} {\bibinfo {author} {\bibfnamefont {K.~V.}\ \bibnamefont
  {Lezhnin}}, \bibinfo {author} {\bibfnamefont {P.~V.}\ \bibnamefont
  {Sasorov}}, \bibinfo {author} {\bibfnamefont {G.}~\bibnamefont {Korn}},\ and\
  \bibinfo {author} {\bibfnamefont {S.~V.}\ \bibnamefont {Bulanov}},\
  }\bibfield  {title} {\bibinfo {title} {{High power gamma flare generation in
  multi-petawatt laser interaction with tailored targets}},\ }\href
  {https://doi.org/10.1063/1.5062849} {\bibfield  {journal} {\bibinfo
  {journal} {Phys. Plasmas}\ }\textbf {\bibinfo {volume} {25}},\ \bibinfo
  {pages} {123105} (\bibinfo {year} {2018})}\BibitemShut {NoStop}%
\bibitem [{\citenamefont {Hadjisolomou}\ \emph {et~al.}(2021)\citenamefont
  {Hadjisolomou}, \citenamefont {Jeong}, \citenamefont {Valenta}, \citenamefont
  {Korn},\ and\ \citenamefont {Bulanov}}]{2021_HadjisolomouP}%
  \BibitemOpen
  \bibfield  {author} {\bibinfo {author} {\bibfnamefont {P.}~\bibnamefont
  {Hadjisolomou}}, \bibinfo {author} {\bibfnamefont {T.~M.}\ \bibnamefont
  {Jeong}}, \bibinfo {author} {\bibfnamefont {P.}~\bibnamefont {Valenta}},
  \bibinfo {author} {\bibfnamefont {G.}~\bibnamefont {Korn}},\ and\ \bibinfo
  {author} {\bibfnamefont {S.~V.}\ \bibnamefont {Bulanov}},\ }\bibfield
  {title} {\bibinfo {title} {{Gamma-ray flash generation in irradiating a thin
  foil target by a single-cycle tightly focused extreme power laser pulse}},\
  }\href {https://doi.org/10.1103/PhysRevE.104.015203} {\bibfield  {journal}
  {\bibinfo  {journal} {Phys. Rev. E}\ }\textbf {\bibinfo {volume} {104}},\
  \bibinfo {pages} {015203} (\bibinfo {year} {2021})}\BibitemShut {NoStop}%
\bibitem [{\citenamefont {Hadjisolomou}\ \emph {et~al.}(2022)\citenamefont
  {Hadjisolomou}, \citenamefont {Jeong}, \citenamefont {Valenta}, \citenamefont
  {Kolenaty}, \citenamefont {Versaci}, \citenamefont {Olšovcová},
  \citenamefont {Ridgers},\ and\ \citenamefont {Bulanov}}]{2022_hadjisolomouP}%
  \BibitemOpen
  \bibfield  {author} {\bibinfo {author} {\bibfnamefont {P.}~\bibnamefont
  {Hadjisolomou}}, \bibinfo {author} {\bibfnamefont {T.~M.}\ \bibnamefont
  {Jeong}}, \bibinfo {author} {\bibfnamefont {P.}~\bibnamefont {Valenta}},
  \bibinfo {author} {\bibfnamefont {D.}~\bibnamefont {Kolenaty}}, \bibinfo
  {author} {\bibfnamefont {R.}~\bibnamefont {Versaci}}, \bibinfo {author}
  {\bibfnamefont {V.}~\bibnamefont {Olšovcová}}, \bibinfo {author}
  {\bibfnamefont {C.~P.}\ \bibnamefont {Ridgers}},\ and\ \bibinfo {author}
  {\bibfnamefont {S.~V.}\ \bibnamefont {Bulanov}},\ }\bibfield  {title}
  {\bibinfo {title} {{Gamma-ray flash in the interaction of a tightly focused
  single-cycle ultraintense laser pulse with a solid target}},\ }\href
  {https://doi.org/10.1017/S0022377821001318} {\bibfield  {journal} {\bibinfo
  {journal} {Journal of Plasma Physics}\ }\textbf {\bibinfo {volume} {88}},\
  \bibinfo {pages} {905880104} (\bibinfo {year} {2022})}\BibitemShut {NoStop}%
\bibitem [{\citenamefont {Mckenna}\ \emph {et~al.}(2016)\citenamefont
  {Mckenna}, \citenamefont {Mangles}, \citenamefont {Sarri},\ and\
  \citenamefont {Schreiber}}]{2016_MckennaP}%
  \BibitemOpen
  \bibfield  {author} {\bibinfo {author} {\bibfnamefont {P.}~\bibnamefont
  {Mckenna}}, \bibinfo {author} {\bibfnamefont {S.~P.~D.}\ \bibnamefont
  {Mangles}}, \bibinfo {author} {\bibfnamefont {G.}~\bibnamefont {Sarri}},\
  and\ \bibinfo {author} {\bibfnamefont {J.}~\bibnamefont {Schreiber}},\
  }\bibfield  {title} {\bibinfo {title} {{High field physics and QED
  experiments at ELI-NP}},\ }\href {http://www.rrp.infim.ro/2016_68_S/S145.pdf}
  {\bibfield  {journal} {\bibinfo  {journal} {Rom Rep Phys}\ }\textbf {\bibinfo
  {volume} {68}},\ \bibinfo {pages} {S145} (\bibinfo {year}
  {2016})}\BibitemShut {NoStop}%
\bibitem [{\citenamefont {Perkins}(2009)}]{2009_PerkinsDH}%
  \BibitemOpen
  \bibfield  {author} {\bibinfo {author} {\bibfnamefont {D.~H.}\ \bibnamefont
  {Perkins}},\ }\href@noop {} {\emph {\bibinfo {title} {{Particle
  astrophysics}}}},\ \bibinfo {edition} {2nd}\ ed.,\ Oxford master series in
  physics, 10\ (\bibinfo  {publisher} {Oxford University Press},\ \bibinfo
  {address} {Oxford},\ \bibinfo {year} {2009})\BibitemShut {NoStop}%
\bibitem [{\citenamefont {Landau}(1944)}]{1944_LandauL}%
  \BibitemOpen
  \bibfield  {author} {\bibinfo {author} {\bibfnamefont {L.}~\bibnamefont
  {Landau}},\ }\bibfield  {title} {\bibinfo {title} {{On the energy loss of
  fast particles by ionization}},\ }\href@noop {} {\bibfield  {journal}
  {\bibinfo  {journal} {J. Phys. USSR}\ }\textbf {\bibinfo {volume} {8}},\
  \bibinfo {pages} {201} (\bibinfo {year} {1944})}\BibitemShut {NoStop}%
\bibitem [{\citenamefont {Bethe}\ and\ \citenamefont
  {Heitler}(1934)}]{1934_BetheH}%
  \BibitemOpen
  \bibfield  {author} {\bibinfo {author} {\bibfnamefont {H.}~\bibnamefont
  {Bethe}}\ and\ \bibinfo {author} {\bibfnamefont {W.}~\bibnamefont
  {Heitler}},\ }\bibfield  {title} {\bibinfo {title} {{On the Stopping of Fast
  Particles and on the Creation of Positive Electrons}},\ }\href
  {https://doi.org/10.1098/rspa.1934.0140} {\bibfield  {journal} {\bibinfo
  {journal} {Proc. R. Soc. Lond. A}\ }\textbf {\bibinfo {volume} {146}},\
  \bibinfo {pages} {83} (\bibinfo {year} {1934})}\BibitemShut {NoStop}%
\bibitem [{\citenamefont {Wheeler}\ and\ \citenamefont
  {Lamb~Jr}(1939)}]{1939_WheelerJA}%
  \BibitemOpen
  \bibfield  {author} {\bibinfo {author} {\bibfnamefont {J.~A.}\ \bibnamefont
  {Wheeler}}\ and\ \bibinfo {author} {\bibfnamefont {W.~E.}\ \bibnamefont
  {Lamb~Jr}},\ }\bibfield  {title} {\bibinfo {title} {{Influence of atomic
  electrons on radiation and pair production}},\ }\href@noop {} {\bibfield
  {journal} {\bibinfo  {journal} {Phys. Rev.}\ }\textbf {\bibinfo {volume}
  {55}},\ \bibinfo {pages} {858} (\bibinfo {year} {1939})}\BibitemShut
  {NoStop}%
\bibitem [{\citenamefont {Koch}\ and\ \citenamefont
  {Motz}(1959)}]{1959_KochHW}%
  \BibitemOpen
  \bibfield  {author} {\bibinfo {author} {\bibfnamefont {H.~W.}\ \bibnamefont
  {Koch}}\ and\ \bibinfo {author} {\bibfnamefont {J.~W.}\ \bibnamefont
  {Motz}},\ }\bibfield  {title} {\bibinfo {title} {{Bremsstrahlung
  cross-section formulas and related data}},\ }\href
  {https://doi.org/10.1103/RevModPhys.31.920} {\bibfield  {journal} {\bibinfo
  {journal} {Rev. Mod. Phys.}\ }\textbf {\bibinfo {volume} {31}},\ \bibinfo
  {pages} {920} (\bibinfo {year} {1959})}\BibitemShut {NoStop}%
\bibitem [{\citenamefont {Aichelin}(1991)}]{1991_AichelinJ}%
  \BibitemOpen
  \bibfield  {author} {\bibinfo {author} {\bibfnamefont {J.}~\bibnamefont
  {Aichelin}},\ }\bibfield  {title} {\bibinfo {title} {{“Quantum” molecular
  dynamics—a dynamical microscopic n-body approach to investigate fragment
  formation and the nuclear equation of state in heavy ion collisions}},\
  }\href {https://doi.org/https://doi.org/10.1016/0370-1573(91)90094-3}
  {\bibfield  {journal} {\bibinfo  {journal} {Phys. Reports}\ }\textbf
  {\bibinfo {volume} {202}},\ \bibinfo {pages} {233} (\bibinfo {year}
  {1991})}\BibitemShut {NoStop}%
\bibitem [{\citenamefont {Compton}(1923)}]{1923_ComptonAH}%
  \BibitemOpen
  \bibfield  {author} {\bibinfo {author} {\bibfnamefont {A.~H.}\ \bibnamefont
  {Compton}},\ }\bibfield  {title} {\bibinfo {title} {{A Quantum theory of the
  scattering of X-rays by light elements}},\ }\href
  {https://doi.org/10.1103/PhysRev.21.483} {\bibfield  {journal} {\bibinfo
  {journal} {Phys. Rev.}\ }\textbf {\bibinfo {volume} {21}},\ \bibinfo {pages}
  {483} (\bibinfo {year} {1923})}\BibitemShut {NoStop}%
\bibitem [{\citenamefont {Hayward}(1970)}]{1970_HaywardE}%
  \BibitemOpen
  \bibfield  {author} {\bibinfo {author} {\bibfnamefont {E.}~\bibnamefont
  {Hayward}},\ }\href@noop {} {\emph {\bibinfo {title} {{Photonuclear
  reactions}}}}\ (\bibinfo  {publisher} {National Bureau of Standards},\
  \bibinfo {address} {Washington, D.C., USA},\ \bibinfo {year}
  {1970})\BibitemShut {NoStop}%
\bibitem [{\citenamefont {Nedorezov}\ \emph {et~al.}(2004)\citenamefont
  {Nedorezov}, \citenamefont {Turinge},\ and\ \citenamefont
  {Shatunov}}]{2004_NedorezovVG}%
  \BibitemOpen
  \bibfield  {author} {\bibinfo {author} {\bibfnamefont {V.~G.}\ \bibnamefont
  {Nedorezov}}, \bibinfo {author} {\bibfnamefont {A.~A.}\ \bibnamefont
  {Turinge}},\ and\ \bibinfo {author} {\bibfnamefont {Y.~M.}\ \bibnamefont
  {Shatunov}},\ }\bibfield  {title} {\bibinfo {title} {{Photonuclear
  experiments with Compton-backscattered gamma beams}},\ }\href
  {https://doi.org/10.1070/pu2004v047n04abeh001743} {\bibfield  {journal}
  {\bibinfo  {journal} {Phys.-Usp.}\ }\textbf {\bibinfo {volume} {47}},\
  \bibinfo {pages} {341} (\bibinfo {year} {2004})}\BibitemShut {NoStop}%
\bibitem [{\citenamefont {Nedorezov}\ \emph {et~al.}(2021)\citenamefont
  {Nedorezov}, \citenamefont {Rykovanov},\ and\ \citenamefont
  {Savel’ev}}]{2021_NedorezovVG}%
  \BibitemOpen
  \bibfield  {author} {\bibinfo {author} {\bibfnamefont {V.~G.}\ \bibnamefont
  {Nedorezov}}, \bibinfo {author} {\bibfnamefont {S.~G.}\ \bibnamefont
  {Rykovanov}},\ and\ \bibinfo {author} {\bibfnamefont {A.~B.}\ \bibnamefont
  {Savel’ev}},\ }\bibfield  {title} {\bibinfo {title} {{Nuclear photonics:
  results and prospects}},\ }\href
  {https://doi.org/10.3367/UFNr.2021.03.038960} {\bibfield  {journal} {\bibinfo
   {journal} {Phys.-Usp.}\ }\textbf {\bibinfo {volume} {191}},\ \bibinfo
  {pages} {1281} (\bibinfo {year} {2021})}\BibitemShut {NoStop}%
\bibitem [{\citenamefont {Budnev}\ \emph {et~al.}(1975)\citenamefont {Budnev},
  \citenamefont {Ginzburg}, \citenamefont {Meledin},\ and\ \citenamefont
  {Serbo}}]{1975_BudnevVM}%
  \BibitemOpen
  \bibfield  {author} {\bibinfo {author} {\bibfnamefont {V.~M.}\ \bibnamefont
  {Budnev}}, \bibinfo {author} {\bibfnamefont {I.~F.}\ \bibnamefont
  {Ginzburg}}, \bibinfo {author} {\bibfnamefont {G.~V.}\ \bibnamefont
  {Meledin}},\ and\ \bibinfo {author} {\bibfnamefont {V.~G.}\ \bibnamefont
  {Serbo}},\ }\bibfield  {title} {\bibinfo {title} {{The two-photon particle
  production mechanism. Physical problems. Applications. Equivalent photon
  approximation}},\ }\href
  {https://doi.org/https://doi.org/10.1016/0370-1573(75)90009-5} {\bibfield
  {journal} {\bibinfo  {journal} {Phys. Reports}\ }\textbf {\bibinfo {volume}
  {15}},\ \bibinfo {pages} {181} (\bibinfo {year} {1975})}\BibitemShut
  {NoStop}%
\bibitem [{\citenamefont {Alejo}\ \emph {et~al.}(2019)\citenamefont {Alejo},
  \citenamefont {Walczak},\ and\ \citenamefont {Sarri}}]{2019_AlejoA}%
  \BibitemOpen
  \bibfield  {author} {\bibinfo {author} {\bibfnamefont {A.}~\bibnamefont
  {Alejo}}, \bibinfo {author} {\bibfnamefont {R.}~\bibnamefont {Walczak}},\
  and\ \bibinfo {author} {\bibfnamefont {G.}~\bibnamefont {Sarri}},\ }\bibfield
   {title} {\bibinfo {title} {{Laser-driven high-quality positron sources as
  possible injectors for plasma-based accelerators}},\ }\href
  {https://doi.org/https://doi.org/10.1038/s41598-019-41650-y} {\bibfield
  {journal} {\bibinfo  {journal} {Sci. Rep.}\ }\textbf {\bibinfo {volume}
  {9}},\ \bibinfo {pages} {1} (\bibinfo {year} {2019})}\BibitemShut {NoStop}%
\bibitem [{\citenamefont {Chen}\ \emph {et~al.}(2010)\citenamefont {Chen},
  \citenamefont {Wilks}, \citenamefont {Meyerhofer}, \citenamefont {Bonlie},
  \citenamefont {Chen}, \citenamefont {Chen}, \citenamefont {Courtois},
  \citenamefont {Elberson}, \citenamefont {Gregori}, \citenamefont {Kruer},
  \citenamefont {Landoas}, \citenamefont {Mithen}, \citenamefont {Myatt},
  \citenamefont {Murphy}, \citenamefont {Nilson}, \citenamefont {Price},
  \citenamefont {Schneider}, \citenamefont {Shepherd}, \citenamefont {Stoeckl},
  \citenamefont {Tabak}, \citenamefont {Tommasini},\ and\ \citenamefont
  {Beiersdorfer}}]{2010_ChenH}%
  \BibitemOpen
  \bibfield  {author} {\bibinfo {author} {\bibfnamefont {H.}~\bibnamefont
  {Chen}}, \bibinfo {author} {\bibfnamefont {S.~C.}\ \bibnamefont {Wilks}},
  \bibinfo {author} {\bibfnamefont {D.~D.}\ \bibnamefont {Meyerhofer}},
  \bibinfo {author} {\bibfnamefont {J.}~\bibnamefont {Bonlie}}, \bibinfo
  {author} {\bibfnamefont {C.~D.}\ \bibnamefont {Chen}}, \bibinfo {author}
  {\bibfnamefont {S.~N.}\ \bibnamefont {Chen}}, \bibinfo {author}
  {\bibfnamefont {C.}~\bibnamefont {Courtois}}, \bibinfo {author}
  {\bibfnamefont {L.}~\bibnamefont {Elberson}}, \bibinfo {author}
  {\bibfnamefont {G.}~\bibnamefont {Gregori}}, \bibinfo {author} {\bibfnamefont
  {W.}~\bibnamefont {Kruer}}, \bibinfo {author} {\bibfnamefont
  {O.}~\bibnamefont {Landoas}}, \bibinfo {author} {\bibfnamefont
  {J.}~\bibnamefont {Mithen}}, \bibinfo {author} {\bibfnamefont
  {J.}~\bibnamefont {Myatt}}, \bibinfo {author} {\bibfnamefont {C.~D.}\
  \bibnamefont {Murphy}}, \bibinfo {author} {\bibfnamefont {P.}~\bibnamefont
  {Nilson}}, \bibinfo {author} {\bibfnamefont {D.}~\bibnamefont {Price}},
  \bibinfo {author} {\bibfnamefont {M.}~\bibnamefont {Schneider}}, \bibinfo
  {author} {\bibfnamefont {R.}~\bibnamefont {Shepherd}}, \bibinfo {author}
  {\bibfnamefont {C.}~\bibnamefont {Stoeckl}}, \bibinfo {author} {\bibfnamefont
  {M.}~\bibnamefont {Tabak}}, \bibinfo {author} {\bibfnamefont
  {R.}~\bibnamefont {Tommasini}},\ and\ \bibinfo {author} {\bibfnamefont
  {P.}~\bibnamefont {Beiersdorfer}},\ }\bibfield  {title} {\bibinfo {title}
  {{Relativistic quasimonoenergetic positron jets from intense laser-solid
  interactions}},\ }\href {https://doi.org/10.1103/PhysRevLett.105.015003}
  {\bibfield  {journal} {\bibinfo  {journal} {Phys. Rev. Lett.}\ }\textbf
  {\bibinfo {volume} {105}},\ \bibinfo {pages} {015003} (\bibinfo {year}
  {2010})}\BibitemShut {NoStop}%
\bibitem [{\citenamefont {Chen}\ \emph {et~al.}(2011)\citenamefont {Chen},
  \citenamefont {Meyerhofer}, \citenamefont {Wilks}, \citenamefont {Cauble},
  \citenamefont {Dollar}, \citenamefont {Falk}, \citenamefont {Gregori},
  \citenamefont {Hazim}, \citenamefont {Moses}, \citenamefont {Murphy},
  \citenamefont {Myatt}, \citenamefont {Park}, \citenamefont {Seely},
  \citenamefont {Shepherd}, \citenamefont {Spitkovsky}, \citenamefont
  {Stoeckl}, \citenamefont {Szabo}, \citenamefont {Tommasini}, \citenamefont
  {Zulick},\ and\ \citenamefont {Beiersdorfer}}]{2011_ChenH}%
  \BibitemOpen
  \bibfield  {author} {\bibinfo {author} {\bibfnamefont {H.}~\bibnamefont
  {Chen}}, \bibinfo {author} {\bibfnamefont {D.~D.}\ \bibnamefont
  {Meyerhofer}}, \bibinfo {author} {\bibfnamefont {S.~C.}\ \bibnamefont
  {Wilks}}, \bibinfo {author} {\bibfnamefont {R.}~\bibnamefont {Cauble}},
  \bibinfo {author} {\bibfnamefont {F.}~\bibnamefont {Dollar}}, \bibinfo
  {author} {\bibfnamefont {K.}~\bibnamefont {Falk}}, \bibinfo {author}
  {\bibfnamefont {G.}~\bibnamefont {Gregori}}, \bibinfo {author} {\bibfnamefont
  {A.}~\bibnamefont {Hazim}}, \bibinfo {author} {\bibfnamefont {E.~I.}\
  \bibnamefont {Moses}}, \bibinfo {author} {\bibfnamefont {C.~D.}\ \bibnamefont
  {Murphy}}, \bibinfo {author} {\bibfnamefont {J.}~\bibnamefont {Myatt}},
  \bibinfo {author} {\bibfnamefont {J.}~\bibnamefont {Park}}, \bibinfo {author}
  {\bibfnamefont {J.}~\bibnamefont {Seely}}, \bibinfo {author} {\bibfnamefont
  {R.}~\bibnamefont {Shepherd}}, \bibinfo {author} {\bibfnamefont
  {A.}~\bibnamefont {Spitkovsky}}, \bibinfo {author} {\bibfnamefont
  {C.}~\bibnamefont {Stoeckl}}, \bibinfo {author} {\bibfnamefont {C.~I.}\
  \bibnamefont {Szabo}}, \bibinfo {author} {\bibfnamefont {R.}~\bibnamefont
  {Tommasini}}, \bibinfo {author} {\bibfnamefont {C.}~\bibnamefont {Zulick}},\
  and\ \bibinfo {author} {\bibfnamefont {P.}~\bibnamefont {Beiersdorfer}},\
  }\bibfield  {title} {\bibinfo {title} {{Towards laboratory produced
  relativistic electron–positron pair plasmas}},\ }\href
  {https://doi.org/https://doi.org/10.1016/j.hedp.2011.05.006} {\bibfield
  {journal} {\bibinfo  {journal} {High Energy Density Phys.}\ }\textbf
  {\bibinfo {volume} {7}},\ \bibinfo {pages} {225} (\bibinfo {year}
  {2011})}\BibitemShut {NoStop}%
\bibitem [{\citenamefont {Chen}\ \emph {et~al.}(2015)\citenamefont {Chen},
  \citenamefont {Fiuza}, \citenamefont {Link}, \citenamefont {Hazi},
  \citenamefont {Hill}, \citenamefont {Hoarty}, \citenamefont {James},
  \citenamefont {Kerr}, \citenamefont {Meyerhofer}, \citenamefont {Myatt},
  \citenamefont {Park}, \citenamefont {Sentoku},\ and\ \citenamefont
  {Williams}}]{2015_ChenH}%
  \BibitemOpen
  \bibfield  {author} {\bibinfo {author} {\bibfnamefont {H.}~\bibnamefont
  {Chen}}, \bibinfo {author} {\bibfnamefont {F.}~\bibnamefont {Fiuza}},
  \bibinfo {author} {\bibfnamefont {A.}~\bibnamefont {Link}}, \bibinfo {author}
  {\bibfnamefont {A.}~\bibnamefont {Hazi}}, \bibinfo {author} {\bibfnamefont
  {M.}~\bibnamefont {Hill}}, \bibinfo {author} {\bibfnamefont {D.}~\bibnamefont
  {Hoarty}}, \bibinfo {author} {\bibfnamefont {S.}~\bibnamefont {James}},
  \bibinfo {author} {\bibfnamefont {S.}~\bibnamefont {Kerr}}, \bibinfo {author}
  {\bibfnamefont {D.~D.}\ \bibnamefont {Meyerhofer}}, \bibinfo {author}
  {\bibfnamefont {J.}~\bibnamefont {Myatt}}, \bibinfo {author} {\bibfnamefont
  {J.}~\bibnamefont {Park}}, \bibinfo {author} {\bibfnamefont {Y.}~\bibnamefont
  {Sentoku}},\ and\ \bibinfo {author} {\bibfnamefont {G.~J.}\ \bibnamefont
  {Williams}},\ }\bibfield  {title} {\bibinfo {title} {{Scaling the yield of
  laser-driven electron-positron jets to laboratory astrophysical
  applications}},\ }\href {https://doi.org/10.1103/PhysRevLett.114.215001}
  {\bibfield  {journal} {\bibinfo  {journal} {Phys. Rev. Lett.}\ }\textbf
  {\bibinfo {volume} {114}},\ \bibinfo {pages} {215001} (\bibinfo {year}
  {2015})}\BibitemShut {NoStop}%
\bibitem [{\citenamefont {Sarri}\ \emph {et~al.}(2015)\citenamefont {Sarri},
  \citenamefont {Poder}, \citenamefont {Cole}, \citenamefont {Schumaker},
  \citenamefont {Di~Piazza}, \citenamefont {Reville}, \citenamefont
  {Dzelzainis}, \citenamefont {Doria}, \citenamefont {Gizzi}, \citenamefont
  {Grittani}, \citenamefont {Kar}, \citenamefont {Keitel}, \citenamefont
  {Krushelnick}, \citenamefont {Kuschel}, \citenamefont {Mangles},
  \citenamefont {Najmudin}, \citenamefont {Shukla}, \citenamefont {Silva},
  \citenamefont {Symes}, \citenamefont {Thomas}, \citenamefont {Vargas},
  \citenamefont {Vieira},\ and\ \citenamefont {Zepf}}]{2015_SarriG}%
  \BibitemOpen
  \bibfield  {author} {\bibinfo {author} {\bibfnamefont {G.}~\bibnamefont
  {Sarri}}, \bibinfo {author} {\bibfnamefont {K.}~\bibnamefont {Poder}},
  \bibinfo {author} {\bibfnamefont {J.~M.}\ \bibnamefont {Cole}}, \bibinfo
  {author} {\bibfnamefont {W.}~\bibnamefont {Schumaker}}, \bibinfo {author}
  {\bibfnamefont {A.}~\bibnamefont {Di~Piazza}}, \bibinfo {author}
  {\bibfnamefont {B.}~\bibnamefont {Reville}}, \bibinfo {author} {\bibfnamefont
  {T.}~\bibnamefont {Dzelzainis}}, \bibinfo {author} {\bibfnamefont
  {D.}~\bibnamefont {Doria}}, \bibinfo {author} {\bibfnamefont {L.~A.}\
  \bibnamefont {Gizzi}}, \bibinfo {author} {\bibfnamefont {G.}~\bibnamefont
  {Grittani}}, \bibinfo {author} {\bibfnamefont {S.}~\bibnamefont {Kar}},
  \bibinfo {author} {\bibfnamefont {C.~H.}\ \bibnamefont {Keitel}}, \bibinfo
  {author} {\bibfnamefont {K.}~\bibnamefont {Krushelnick}}, \bibinfo {author}
  {\bibfnamefont {S.}~\bibnamefont {Kuschel}}, \bibinfo {author} {\bibfnamefont
  {S.~P.~D.}\ \bibnamefont {Mangles}}, \bibinfo {author} {\bibfnamefont
  {Z.}~\bibnamefont {Najmudin}}, \bibinfo {author} {\bibfnamefont
  {N.}~\bibnamefont {Shukla}}, \bibinfo {author} {\bibfnamefont {L.~O.}\
  \bibnamefont {Silva}}, \bibinfo {author} {\bibfnamefont {D.}~\bibnamefont
  {Symes}}, \bibinfo {author} {\bibfnamefont {A.~G.~R.}\ \bibnamefont
  {Thomas}}, \bibinfo {author} {\bibfnamefont {M.}~\bibnamefont {Vargas}},
  \bibinfo {author} {\bibfnamefont {J.}~\bibnamefont {Vieira}},\ and\ \bibinfo
  {author} {\bibfnamefont {M.}~\bibnamefont {Zepf}},\ }\bibfield  {title}
  {\bibinfo {title} {{Generation of neutral and high-density electron--positron
  pair plasmas in the laboratory}},\ }\href
  {https://doi.org/https://doi.org/10.1038/ncomms7747} {\bibfield  {journal}
  {\bibinfo  {journal} {Nat. Commun.}\ }\textbf {\bibinfo {volume} {6}},\
  \bibinfo {pages} {1} (\bibinfo {year} {2015})}\BibitemShut {NoStop}%
\bibitem [{\citenamefont {Xu}\ \emph {et~al.}(2016)\citenamefont {Xu},
  \citenamefont {Shen}, \citenamefont {Xu}, \citenamefont {Li}, \citenamefont
  {Yu}, \citenamefont {Li}, \citenamefont {Lu}, \citenamefont {Wang},
  \citenamefont {Wang}, \citenamefont {Liang}, \citenamefont {Leng},
  \citenamefont {Li},\ and\ \citenamefont {Xu}}]{2016_XuT}%
  \BibitemOpen
  \bibfield  {author} {\bibinfo {author} {\bibfnamefont {T.}~\bibnamefont
  {Xu}}, \bibinfo {author} {\bibfnamefont {B.}~\bibnamefont {Shen}}, \bibinfo
  {author} {\bibfnamefont {J.}~\bibnamefont {Xu}}, \bibinfo {author}
  {\bibfnamefont {S.}~\bibnamefont {Li}}, \bibinfo {author} {\bibfnamefont
  {Y.}~\bibnamefont {Yu}}, \bibinfo {author} {\bibfnamefont {J.}~\bibnamefont
  {Li}}, \bibinfo {author} {\bibfnamefont {X.}~\bibnamefont {Lu}}, \bibinfo
  {author} {\bibfnamefont {C.}~\bibnamefont {Wang}}, \bibinfo {author}
  {\bibfnamefont {X.}~\bibnamefont {Wang}}, \bibinfo {author} {\bibfnamefont
  {X.}~\bibnamefont {Liang}}, \bibinfo {author} {\bibfnamefont
  {Y.}~\bibnamefont {Leng}}, \bibinfo {author} {\bibfnamefont {R.}~\bibnamefont
  {Li}},\ and\ \bibinfo {author} {\bibfnamefont {Z.}~\bibnamefont {Xu}},\
  }\bibfield  {title} {\bibinfo {title} {{Ultrashort megaelectronvolt positron
  beam generation based on laser-accelerated electrons}},\ }\href
  {https://doi.org/https://doi.org/10.1063/1.4943280} {\bibfield  {journal}
  {\bibinfo  {journal} {Phys. Plasmas}\ }\textbf {\bibinfo {volume} {23}},\
  \bibinfo {pages} {033109} (\bibinfo {year} {2016})}\BibitemShut {NoStop}%
\bibitem [{\citenamefont {Audet}\ \emph {et~al.}(2021)\citenamefont {Audet},
  \citenamefont {Alejo}, \citenamefont {Calvin}, \citenamefont {Cunningham},
  \citenamefont {Frazer}, \citenamefont {Nersisyan}, \citenamefont {Phipps},
  \citenamefont {Warwick}, \citenamefont {Sarri}, \citenamefont {Hafz},
  \citenamefont {Kamperidis}, \citenamefont {Li},\ and\ \citenamefont
  {Papp}}]{2021_AudetTL}%
  \BibitemOpen
  \bibfield  {author} {\bibinfo {author} {\bibfnamefont {T.~L.}\ \bibnamefont
  {Audet}}, \bibinfo {author} {\bibfnamefont {A.}~\bibnamefont {Alejo}},
  \bibinfo {author} {\bibfnamefont {L.}~\bibnamefont {Calvin}}, \bibinfo
  {author} {\bibfnamefont {M.~H.}\ \bibnamefont {Cunningham}}, \bibinfo
  {author} {\bibfnamefont {G.~R.}\ \bibnamefont {Frazer}}, \bibinfo {author}
  {\bibfnamefont {G.}~\bibnamefont {Nersisyan}}, \bibinfo {author}
  {\bibfnamefont {M.}~\bibnamefont {Phipps}}, \bibinfo {author} {\bibfnamefont
  {J.~R.}\ \bibnamefont {Warwick}}, \bibinfo {author} {\bibfnamefont
  {G.}~\bibnamefont {Sarri}}, \bibinfo {author} {\bibfnamefont {N.~A.~M.}\
  \bibnamefont {Hafz}}, \bibinfo {author} {\bibfnamefont {C.}~\bibnamefont
  {Kamperidis}}, \bibinfo {author} {\bibfnamefont {S.}~\bibnamefont {Li}},\
  and\ \bibinfo {author} {\bibfnamefont {D.}~\bibnamefont {Papp}},\ }\bibfield
  {title} {\bibinfo {title} {{Ultrashort, MeV-scale laser-plasma positron
  source for positron annihilation lifetime spectroscopy}},\ }\href
  {https://doi.org/10.1103/PhysRevAccelBeams.24.073402} {\bibfield  {journal}
  {\bibinfo  {journal} {Phys. Rev. Accel. Beams}\ }\textbf {\bibinfo {volume}
  {24}},\ \bibinfo {pages} {073402} (\bibinfo {year} {2021})}\BibitemShut
  {NoStop}%
\bibitem [{\citenamefont {Sarri}(2021)}]{2021_SarriG}%
  \BibitemOpen
  \bibfield  {author} {\bibinfo {author} {\bibfnamefont {G.}~\bibnamefont
  {Sarri}},\ }\bibfield  {title} {\bibinfo {title} {{Laser-driven positron
  sources for applications in fundamental science and industry}},\ }in\ \href
  {https://doi.org/https://doi.org/10.1117/12.2596510} {\emph {\bibinfo
  {booktitle} {Applying Laser-driven Particle Acceleration II, Medical and
  Nonmedical Uses of Distinctive Energetic Particle and Photon Sources: SPIE
  Optics + Optoelectronics Industry Event}}},\ Vol.\ \bibinfo {volume}
  {11790},\ \bibinfo {editor} {edited by\ \bibinfo {editor} {\bibfnamefont
  {P.~R.}\ \bibnamefont {Bolton}}},\ \bibinfo {organization} {International
  Society for Optics and Photonics}\ (\bibinfo  {publisher} {SPIE},\ \bibinfo
  {year} {2021})\BibitemShut {NoStop}%
\bibitem [{\citenamefont {Higginson}\ \emph {et~al.}(2010)\citenamefont
  {Higginson}, \citenamefont {McNaney}, \citenamefont {Swift}, \citenamefont
  {Bartal}, \citenamefont {Hey}, \citenamefont {Kodama}, \citenamefont
  {Le~Pape}, \citenamefont {Mackinnon}, \citenamefont {Mariscal}, \citenamefont
  {Nakamura}, \citenamefont {Nakanii}, \citenamefont {Tanaka},\ and\
  \citenamefont {Beg}}]{2010_HigginsonDP}%
  \BibitemOpen
  \bibfield  {author} {\bibinfo {author} {\bibfnamefont {D.~P.}\ \bibnamefont
  {Higginson}}, \bibinfo {author} {\bibfnamefont {J.~M.}\ \bibnamefont
  {McNaney}}, \bibinfo {author} {\bibfnamefont {D.~C.}\ \bibnamefont {Swift}},
  \bibinfo {author} {\bibfnamefont {T.}~\bibnamefont {Bartal}}, \bibinfo
  {author} {\bibfnamefont {D.~S.}\ \bibnamefont {Hey}}, \bibinfo {author}
  {\bibfnamefont {R.}~\bibnamefont {Kodama}}, \bibinfo {author} {\bibfnamefont
  {S.}~\bibnamefont {Le~Pape}}, \bibinfo {author} {\bibfnamefont
  {A.}~\bibnamefont {Mackinnon}}, \bibinfo {author} {\bibfnamefont
  {D.}~\bibnamefont {Mariscal}}, \bibinfo {author} {\bibfnamefont
  {H.}~\bibnamefont {Nakamura}}, \bibinfo {author} {\bibfnamefont
  {N.}~\bibnamefont {Nakanii}}, \bibinfo {author} {\bibfnamefont {K.~A.}\
  \bibnamefont {Tanaka}},\ and\ \bibinfo {author} {\bibfnamefont {F.~N.}\
  \bibnamefont {Beg}},\ }\bibfield  {title} {\bibinfo {title} {{Laser generated
  neutron source for neutron resonance spectroscopy}},\ }\href
  {https://doi.org/https://doi.org/10.1063/1.3484218} {\bibfield  {journal}
  {\bibinfo  {journal} {Phys. Plasmas}\ }\textbf {\bibinfo {volume} {17}},\
  \bibinfo {pages} {100701} (\bibinfo {year} {2010})}\BibitemShut {NoStop}%
\bibitem [{\citenamefont {Zimmer}\ \emph {et~al.}(2020)\citenamefont {Zimmer},
  \citenamefont {Scheuren}, \citenamefont {Kleinschmidt}, \citenamefont
  {Tebartz}, \citenamefont {Ebert}, \citenamefont {Ding}, \citenamefont
  {Hartnagel},\ and\ \citenamefont {Roth}}]{2020_ZimmerM}%
  \BibitemOpen
  \bibfield  {author} {\bibinfo {author} {\bibfnamefont {M.}~\bibnamefont
  {Zimmer}}, \bibinfo {author} {\bibfnamefont {S.}~\bibnamefont {Scheuren}},
  \bibinfo {author} {\bibfnamefont {A.}~\bibnamefont {Kleinschmidt}}, \bibinfo
  {author} {\bibfnamefont {A.}~\bibnamefont {Tebartz}}, \bibinfo {author}
  {\bibfnamefont {T.}~\bibnamefont {Ebert}}, \bibinfo {author} {\bibfnamefont
  {J.}~\bibnamefont {Ding}}, \bibinfo {author} {\bibfnamefont {D.}~\bibnamefont
  {Hartnagel}},\ and\ \bibinfo {author} {\bibfnamefont {M.}~\bibnamefont
  {Roth}},\ }\bibfield  {title} {\bibinfo {title} {{Development of a setup for
  material identification based on laser-driven neutron resonance
  spectroscopy}},\ }in\ \href {https://doi.org/10.1051/epjconf/202023101006}
  {\emph {\bibinfo {booktitle} {EPJ Web of Conferences}}},\ Vol.\ \bibinfo
  {volume} {231}\ (\bibinfo {organization} {EDP Sciences},\ \bibinfo {year}
  {2020})\ p.\ \bibinfo {pages} {01006}\BibitemShut {NoStop}%
\bibitem [{\citenamefont {Aksenov}\ and\ \citenamefont
  {Balagurov}(2016)}]{2016_AksenovVL}%
  \BibitemOpen
  \bibfield  {author} {\bibinfo {author} {\bibfnamefont {V.~L.}\ \bibnamefont
  {Aksenov}}\ and\ \bibinfo {author} {\bibfnamefont {A.~M.}\ \bibnamefont
  {Balagurov}},\ }\bibfield  {title} {\bibinfo {title} {{Neutron diffraction on
  pulsed sources}},\ }\href {https://doi.org/10.3367/ufne.0186.201603e.0293}
  {\bibfield  {journal} {\bibinfo  {journal} {Phys.-Usp.}\ }\textbf {\bibinfo
  {volume} {59}},\ \bibinfo {pages} {279} (\bibinfo {year} {2016})}\BibitemShut
  {NoStop}%
\bibitem [{\citenamefont {Vogel}\ \emph {et~al.}(2020)\citenamefont {Vogel},
  \citenamefont {Fernandez}, \citenamefont {Gautier}, \citenamefont {Mitura},
  \citenamefont {Roth},\ and\ \citenamefont {Schoenberg}}]{2020_VogelS}%
  \BibitemOpen
  \bibfield  {author} {\bibinfo {author} {\bibfnamefont {S.~C.}\ \bibnamefont
  {Vogel}}, \bibinfo {author} {\bibfnamefont {J.~C.}\ \bibnamefont
  {Fernandez}}, \bibinfo {author} {\bibfnamefont {D.~C.}\ \bibnamefont
  {Gautier}}, \bibinfo {author} {\bibfnamefont {N.}~\bibnamefont {Mitura}},
  \bibinfo {author} {\bibfnamefont {M.}~\bibnamefont {Roth}},\ and\ \bibinfo
  {author} {\bibfnamefont {K.~F.}\ \bibnamefont {Schoenberg}},\ }\bibfield
  {title} {\bibinfo {title} {{Short-pulse laser-driven moderated neutron
  source}},\ }in\ \href {https://doi.org/10.1051/epjconf/202023101008} {\emph
  {\bibinfo {booktitle} {EPJ Web of Conferences}}},\ Vol.\ \bibinfo {volume}
  {231}\ (\bibinfo {organization} {EDP Sciences},\ \bibinfo {year} {2020})\ p.\
  \bibinfo {pages} {01008}\BibitemShut {NoStop}%
\bibitem [{\citenamefont {Revol}(2001)}]{2001_RevolJP}%
  \BibitemOpen
  \bibfield  {author} {\bibinfo {author} {\bibfnamefont {J.~P.}\ \bibnamefont
  {Revol}},\ }\bibfield  {title} {\bibinfo {title} {{An accelerator-driven
  system for the destruction of nuclear waste}},\ }\href
  {https://doi.org/https://doi.org/10.1016/S0149-1970(00)00100-1} {\bibfield
  {journal} {\bibinfo  {journal} {Prog. Nucl. Energy}\ }\textbf {\bibinfo
  {volume} {38}},\ \bibinfo {pages} {153} (\bibinfo {year} {2001})}\BibitemShut
  {NoStop}%
\bibitem [{\citenamefont {von Lensa}\ \emph {et~al.}(2008)\citenamefont {von
  Lensa}, \citenamefont {Nabbi},\ and\ \citenamefont {Rossbach}}]{2008_LensaW}%
  \BibitemOpen
  \bibfield  {author} {\bibinfo {author} {\bibfnamefont {W.}~\bibnamefont {von
  Lensa}}, \bibinfo {author} {\bibfnamefont {R.}~\bibnamefont {Nabbi}},\ and\
  \bibinfo {author} {\bibfnamefont {M.}~\bibnamefont {Rossbach}},\ }\href@noop
  {} {\bibinfo {title} {{RED-IMPACT. Impact of partitioning, transmutation and
  waste reduction technologies on the final nuclear waste disposal. Synthesis
  report}}} (\bibinfo {year} {2008})\BibitemShut {NoStop}%
\bibitem [{\citenamefont {Nishiuchi}\ \emph {et~al.}(2016)\citenamefont
  {Nishiuchi}, \citenamefont {Sakaki}, \citenamefont {Esirkepov}, \citenamefont
  {Nishio}, \citenamefont {Pikuz}, \citenamefont {Faenov}, \citenamefont
  {Skobelev}, \citenamefont {Orlandi}, \citenamefont {Pirozhkov}, \citenamefont
  {Sagisaka} \emph {et~al.}}]{2016_NishiuchiM}%
  \BibitemOpen
  \bibfield  {author} {\bibinfo {author} {\bibfnamefont {M.}~\bibnamefont
  {Nishiuchi}}, \bibinfo {author} {\bibfnamefont {H.}~\bibnamefont {Sakaki}},
  \bibinfo {author} {\bibfnamefont {T.~Z.}\ \bibnamefont {Esirkepov}}, \bibinfo
  {author} {\bibfnamefont {K.}~\bibnamefont {Nishio}}, \bibinfo {author}
  {\bibfnamefont {T.~A.}\ \bibnamefont {Pikuz}}, \bibinfo {author}
  {\bibfnamefont {A.~Y.}\ \bibnamefont {Faenov}}, \bibinfo {author}
  {\bibfnamefont {I.~Y.}\ \bibnamefont {Skobelev}}, \bibinfo {author}
  {\bibfnamefont {R.}~\bibnamefont {Orlandi}}, \bibinfo {author} {\bibfnamefont
  {A.~S.}\ \bibnamefont {Pirozhkov}}, \bibinfo {author} {\bibfnamefont
  {A.}~\bibnamefont {Sagisaka}}, \emph {et~al.},\ }\bibfield  {title} {\bibinfo
  {title} {{Towards a novel laser-driven method of exotic nuclei extraction-
  acceleration for fundamental physics and technology}},\ }\href
  {https://doi.org/10.1134/S1063780X1604005X} {\bibfield  {journal} {\bibinfo
  {journal} {Plasma Phys. Rep.}\ }\textbf {\bibinfo {volume} {42}},\ \bibinfo
  {pages} {327} (\bibinfo {year} {2016})}\BibitemShut {NoStop}%
\bibitem [{\citenamefont {Sun}(2021)}]{2021_SunZ}%
  \BibitemOpen
  \bibfield  {author} {\bibinfo {author} {\bibfnamefont {Z.}~\bibnamefont
  {Sun}},\ }\bibfield  {title} {\bibinfo {title} {{Review: Production of
  nuclear medicine radioisotopes with ultra-intense lasers}},\ }\href
  {https://doi.org/10.1063/5.0042796} {\bibfield  {journal} {\bibinfo
  {journal} {AIP Adv.}\ }\textbf {\bibinfo {volume} {11}},\ \bibinfo {pages}
  {040701} (\bibinfo {year} {2021})}\BibitemShut {NoStop}%
\bibitem [{\citenamefont {Arber}\ \emph {et~al.}(2015)\citenamefont {Arber},
  \citenamefont {Bennett}, \citenamefont {Brady}, \citenamefont
  {Lawrence-Douglas}, \citenamefont {Ramsay}, \citenamefont {Sircombe},
  \citenamefont {Gillies}, \citenamefont {Evans}, \citenamefont {Schmitz},
  \citenamefont {Bell},\ and\ \citenamefont {Ridgers}}]{2015_ArberTD}%
  \BibitemOpen
  \bibfield  {author} {\bibinfo {author} {\bibfnamefont {T.~D.}\ \bibnamefont
  {Arber}}, \bibinfo {author} {\bibfnamefont {K.}~\bibnamefont {Bennett}},
  \bibinfo {author} {\bibfnamefont {C.~S.}\ \bibnamefont {Brady}}, \bibinfo
  {author} {\bibfnamefont {A.}~\bibnamefont {Lawrence-Douglas}}, \bibinfo
  {author} {\bibfnamefont {M.~G.}\ \bibnamefont {Ramsay}}, \bibinfo {author}
  {\bibfnamefont {N.~J.}\ \bibnamefont {Sircombe}}, \bibinfo {author}
  {\bibfnamefont {P.}~\bibnamefont {Gillies}}, \bibinfo {author} {\bibfnamefont
  {R.~G.}\ \bibnamefont {Evans}}, \bibinfo {author} {\bibfnamefont
  {H.}~\bibnamefont {Schmitz}}, \bibinfo {author} {\bibfnamefont {A.~R.}\
  \bibnamefont {Bell}},\ and\ \bibinfo {author} {\bibfnamefont {C.~P.}\
  \bibnamefont {Ridgers}},\ }\bibfield  {title} {\bibinfo {title}
  {{Contemporary particle-in-cell approach to laser-plasma modelling}},\ }\href
  {https://doi.org/10.1088/0741-3335/57/11/113001} {\bibfield  {journal}
  {\bibinfo  {journal} {Plasma Phys. Control. Fusion}\ }\textbf {\bibinfo
  {volume} {57}},\ \bibinfo {pages} {113001} (\bibinfo {year}
  {2015})}\BibitemShut {NoStop}%
\bibitem [{\citenamefont {Jeong}\ \emph {et~al.}(2015)\citenamefont {Jeong},
  \citenamefont {Weber}, \citenamefont {Le~Garrec}, \citenamefont {Margarone},
  \citenamefont {Mocek},\ and\ \citenamefont {Korn}}]{2015_JeongTM}%
  \BibitemOpen
  \bibfield  {author} {\bibinfo {author} {\bibfnamefont {T.~M.}\ \bibnamefont
  {Jeong}}, \bibinfo {author} {\bibfnamefont {S.}~\bibnamefont {Weber}},
  \bibinfo {author} {\bibfnamefont {B.}~\bibnamefont {Le~Garrec}}, \bibinfo
  {author} {\bibfnamefont {D.}~\bibnamefont {Margarone}}, \bibinfo {author}
  {\bibfnamefont {T.}~\bibnamefont {Mocek}},\ and\ \bibinfo {author}
  {\bibfnamefont {G.}~\bibnamefont {Korn}},\ }\bibfield  {title} {\bibinfo
  {title} {{Spatio-temporal modification of femtosecond focal spot under tight
  focusing condition}},\ }\href {https://doi.org/10.1364/OE.23.011641}
  {\bibfield  {journal} {\bibinfo  {journal} {Opt. Express}\ }\textbf {\bibinfo
  {volume} {23}},\ \bibinfo {pages} {11641} (\bibinfo {year}
  {2015})}\BibitemShut {NoStop}%
\bibitem [{\citenamefont {Jeong}\ \emph {et~al.}(2018)\citenamefont {Jeong},
  \citenamefont {Bulanov}, \citenamefont {Weber},\ and\ \citenamefont
  {Korn}}]{2018_JeongTM}%
  \BibitemOpen
  \bibfield  {author} {\bibinfo {author} {\bibfnamefont {T.~M.}\ \bibnamefont
  {Jeong}}, \bibinfo {author} {\bibfnamefont {S.}~\bibnamefont {Bulanov}},
  \bibinfo {author} {\bibfnamefont {S.}~\bibnamefont {Weber}},\ and\ \bibinfo
  {author} {\bibfnamefont {G.}~\bibnamefont {Korn}},\ }\bibfield  {title}
  {\bibinfo {title} {{Analysis on the longitudinal field strength formed by
  tightly-focused radially-polarized femtosecond petawatt laser pulse}},\
  }\href {https://doi.org/10.1364/OE.26.033091} {\bibfield  {journal} {\bibinfo
   {journal} {Opt. Express}\ }\textbf {\bibinfo {volume} {26}},\ \bibinfo
  {pages} {33091} (\bibinfo {year} {2018})}\BibitemShut {NoStop}%
\bibitem [{\citenamefont {Ahdida}\ \emph {et~al.}(2022)\citenamefont {Ahdida},
  \citenamefont {Bozzato}, \citenamefont {Calzolari}, \citenamefont {Cerutti},
  \citenamefont {Charitonidis}, \citenamefont {Cimmino}, \citenamefont
  {Coronetti}, \citenamefont {D'alessandro}, \citenamefont {Donadon~Servelle},
  \citenamefont {Esposito}, \citenamefont {Froeschl}, \citenamefont {Alía},
  \citenamefont {Gerbershagen}, \citenamefont {Gilardoni}, \citenamefont
  {Horváth}, \citenamefont {Hugo}, \citenamefont {Infantino}, \citenamefont
  {Kouskoura}, \citenamefont {Lechner}, \citenamefont {Lefebvre}, \citenamefont
  {Lerner}, \citenamefont {Magistris}, \citenamefont {Manousos}, \citenamefont
  {Moryc}, \citenamefont {Ruiz}, \citenamefont {Pozzi}, \citenamefont
  {Prelipcean}, \citenamefont {Roesler}, \citenamefont {Rossi}, \citenamefont
  {Sabate~Gilarte}, \citenamefont {Salvat~Pujol}, \citenamefont {Schoofs},
  \citenamefont {Stránský}, \citenamefont {Theis}, \citenamefont {Tsinganis},
  \citenamefont {Versaci}, \citenamefont {Vlachoudis}, \citenamefont {Waets},\
  and\ \citenamefont {Widorski}}]{AhdidaC}%
  \BibitemOpen
  \bibfield  {author} {\bibinfo {author} {\bibfnamefont {C.}~\bibnamefont
  {Ahdida}}, \bibinfo {author} {\bibfnamefont {D.}~\bibnamefont {Bozzato}},
  \bibinfo {author} {\bibfnamefont {D.}~\bibnamefont {Calzolari}}, \bibinfo
  {author} {\bibfnamefont {F.}~\bibnamefont {Cerutti}}, \bibinfo {author}
  {\bibfnamefont {N.}~\bibnamefont {Charitonidis}}, \bibinfo {author}
  {\bibfnamefont {A.}~\bibnamefont {Cimmino}}, \bibinfo {author} {\bibfnamefont
  {A.}~\bibnamefont {Coronetti}}, \bibinfo {author} {\bibfnamefont
  {G.}~\bibnamefont {D'alessandro}}, \bibinfo {author} {\bibfnamefont
  {A.}~\bibnamefont {Donadon~Servelle}}, \bibinfo {author} {\bibfnamefont
  {L.}~\bibnamefont {Esposito}}, \bibinfo {author} {\bibfnamefont
  {R.}~\bibnamefont {Froeschl}}, \bibinfo {author} {\bibfnamefont {R.~G.}\
  \bibnamefont {Alía}}, \bibinfo {author} {\bibfnamefont {A.}~\bibnamefont
  {Gerbershagen}}, \bibinfo {author} {\bibfnamefont {S.}~\bibnamefont
  {Gilardoni}}, \bibinfo {author} {\bibfnamefont {D.}~\bibnamefont {Horváth}},
  \bibinfo {author} {\bibfnamefont {G.}~\bibnamefont {Hugo}}, \bibinfo {author}
  {\bibfnamefont {A.}~\bibnamefont {Infantino}}, \bibinfo {author}
  {\bibfnamefont {V.}~\bibnamefont {Kouskoura}}, \bibinfo {author}
  {\bibfnamefont {A.}~\bibnamefont {Lechner}}, \bibinfo {author} {\bibfnamefont
  {B.}~\bibnamefont {Lefebvre}}, \bibinfo {author} {\bibfnamefont
  {G.}~\bibnamefont {Lerner}}, \bibinfo {author} {\bibfnamefont
  {M.}~\bibnamefont {Magistris}}, \bibinfo {author} {\bibfnamefont
  {A.}~\bibnamefont {Manousos}}, \bibinfo {author} {\bibfnamefont
  {G.}~\bibnamefont {Moryc}}, \bibinfo {author} {\bibfnamefont {F.~O.}\
  \bibnamefont {Ruiz}}, \bibinfo {author} {\bibfnamefont {F.}~\bibnamefont
  {Pozzi}}, \bibinfo {author} {\bibfnamefont {D.}~\bibnamefont {Prelipcean}},
  \bibinfo {author} {\bibfnamefont {S.}~\bibnamefont {Roesler}}, \bibinfo
  {author} {\bibfnamefont {R.}~\bibnamefont {Rossi}}, \bibinfo {author}
  {\bibfnamefont {M.}~\bibnamefont {Sabate~Gilarte}}, \bibinfo {author}
  {\bibfnamefont {F.}~\bibnamefont {Salvat~Pujol}}, \bibinfo {author}
  {\bibfnamefont {P.}~\bibnamefont {Schoofs}}, \bibinfo {author} {\bibfnamefont
  {V.}~\bibnamefont {Stránský}}, \bibinfo {author} {\bibfnamefont
  {C.}~\bibnamefont {Theis}}, \bibinfo {author} {\bibfnamefont
  {A.}~\bibnamefont {Tsinganis}}, \bibinfo {author} {\bibfnamefont
  {R.}~\bibnamefont {Versaci}}, \bibinfo {author} {\bibfnamefont
  {V.}~\bibnamefont {Vlachoudis}}, \bibinfo {author} {\bibfnamefont
  {A.}~\bibnamefont {Waets}},\ and\ \bibinfo {author} {\bibfnamefont
  {M.}~\bibnamefont {Widorski}},\ }\bibfield  {title} {\bibinfo {title} {{New
  capabilities of the FLUKA multi-purpose code}},\ }\href
  {https://doi.org/10.3389/fphy.2021.788253} {\bibfield  {journal} {\bibinfo
  {journal} {Front. Phys.}\ }\textbf {\bibinfo {volume} {9}},\ \bibinfo {pages}
  {788253} (\bibinfo {year} {2022})}\BibitemShut {NoStop}%
\bibitem [{\citenamefont {Battistoni}\ \emph {et~al.}(2015)\citenamefont
  {Battistoni}, \citenamefont {Boehlen}, \citenamefont {Cerutti}, \citenamefont
  {Chin}, \citenamefont {Esposito}, \citenamefont {Fassò}, \citenamefont
  {Ferrari}, \citenamefont {Lechner}, \citenamefont {Empl}, \citenamefont
  {Mairani}, \citenamefont {Mereghetti}, \citenamefont {Ortega}, \citenamefont
  {Ranft}, \citenamefont {Roesler}, \citenamefont {Sala}, \citenamefont
  {Vlachoudis},\ and\ \citenamefont {Smirnov}}]{2015_BattistoniG}%
  \BibitemOpen
  \bibfield  {author} {\bibinfo {author} {\bibfnamefont {G.}~\bibnamefont
  {Battistoni}}, \bibinfo {author} {\bibfnamefont {T.}~\bibnamefont {Boehlen}},
  \bibinfo {author} {\bibfnamefont {F.}~\bibnamefont {Cerutti}}, \bibinfo
  {author} {\bibfnamefont {P.~W.}\ \bibnamefont {Chin}}, \bibinfo {author}
  {\bibfnamefont {L.~S.}\ \bibnamefont {Esposito}}, \bibinfo {author}
  {\bibfnamefont {A.}~\bibnamefont {Fassò}}, \bibinfo {author} {\bibfnamefont
  {A.}~\bibnamefont {Ferrari}}, \bibinfo {author} {\bibfnamefont
  {A.}~\bibnamefont {Lechner}}, \bibinfo {author} {\bibfnamefont
  {A.}~\bibnamefont {Empl}}, \bibinfo {author} {\bibfnamefont {A.}~\bibnamefont
  {Mairani}}, \bibinfo {author} {\bibfnamefont {A.}~\bibnamefont {Mereghetti}},
  \bibinfo {author} {\bibfnamefont {P.~G.}\ \bibnamefont {Ortega}}, \bibinfo
  {author} {\bibfnamefont {J.}~\bibnamefont {Ranft}}, \bibinfo {author}
  {\bibfnamefont {S.}~\bibnamefont {Roesler}}, \bibinfo {author} {\bibfnamefont
  {P.~R.}\ \bibnamefont {Sala}}, \bibinfo {author} {\bibfnamefont
  {V.}~\bibnamefont {Vlachoudis}},\ and\ \bibinfo {author} {\bibfnamefont
  {G.}~\bibnamefont {Smirnov}},\ }\bibfield  {title} {\bibinfo {title}
  {{Overview of the FLUKA code}},\ }\href
  {https://doi.org/https://doi.org/10.1016/j.anucene.2014.11.007} {\bibfield
  {journal} {\bibinfo  {journal} {Ann. Nucl. Energy}\ }\textbf {\bibinfo
  {volume} {82}},\ \bibinfo {pages} {10} (\bibinfo {year} {2015})}\BibitemShut
  {NoStop}%
\bibitem [{\citenamefont {Vlachoudis}(2009)}]{2009_VlachoudisV}%
  \BibitemOpen
  \bibfield  {author} {\bibinfo {author} {\bibfnamefont {V.}~\bibnamefont
  {Vlachoudis}},\ }\bibfield  {title} {\bibinfo {title} {{FLAIR: a powerful but
  user friendly graphical interface for FLUKA}},\ }in\ \href@noop {} {\emph
  {\bibinfo {booktitle} {Proc. Int. Conf. on Mathematics, Computational Methods
  \& Reactor Physics (M\&C 2009), Saratoga Springs, New York}}},\ Vol.\
  \bibinfo {volume} {176}\ (\bibinfo {year} {2009})\BibitemShut {NoStop}%
\bibitem [{\citenamefont {Ferrari}\ \emph {et~al.}(2005)\citenamefont
  {Ferrari}, \citenamefont {Sala}, \citenamefont {Fasso}, \citenamefont
  {Ranft},\ and\ \citenamefont {Siegen}}]{2005_FerrariA}%
  \BibitemOpen
  \bibfield  {author} {\bibinfo {author} {\bibfnamefont {A.}~\bibnamefont
  {Ferrari}}, \bibinfo {author} {\bibfnamefont {P.~R.}\ \bibnamefont {Sala}},
  \bibinfo {author} {\bibfnamefont {A.}~\bibnamefont {Fasso}}, \bibinfo
  {author} {\bibfnamefont {J.}~\bibnamefont {Ranft}},\ and\ \bibinfo {author}
  {\bibfnamefont {U.}~\bibnamefont {Siegen}},\ }\href
  {https://doi.org/10.2172/877507} {\emph {\bibinfo {title} {{FLUKA: a
  multi-particle transport code}}}},\ \bibinfo {type} {Tech. Rep.}\ \bibinfo
  {number} {CERN-2005-010, SLAC-R-773, INFN-TC-05-11, CERN-2005-10}\ (\bibinfo
  {institution} {SLAC National Accelerator Lab.},\ \bibinfo {address} {Menlo
  Park, CA (United States)},\ \bibinfo {year} {2005})\BibitemShut {NoStop}%
\bibitem [{\citenamefont {Andersen}\ \emph {et~al.}(2004)\citenamefont
  {Andersen}, \citenamefont {Ballarini}, \citenamefont {Battistoni},
  \citenamefont {Campanella}, \citenamefont {Carboni}, \citenamefont {Cerutti},
  \citenamefont {Empl}, \citenamefont {Fassò}, \citenamefont {Ferrari},
  \citenamefont {E.}, \citenamefont {Garzelli}, \citenamefont {Lee},
  \citenamefont {Ottolenghi}, \citenamefont {Pelliccioni}, \citenamefont
  {Pinsky}, \citenamefont {Ranft}, \citenamefont {Roesler}, \citenamefont
  {Sala},\ and\ \citenamefont {Wilson}}]{2004_AndersenV}%
  \BibitemOpen
  \bibfield  {author} {\bibinfo {author} {\bibfnamefont {V.}~\bibnamefont
  {Andersen}}, \bibinfo {author} {\bibfnamefont {F.}~\bibnamefont {Ballarini}},
  \bibinfo {author} {\bibfnamefont {G.}~\bibnamefont {Battistoni}}, \bibinfo
  {author} {\bibfnamefont {M.}~\bibnamefont {Campanella}}, \bibinfo {author}
  {\bibfnamefont {M.}~\bibnamefont {Carboni}}, \bibinfo {author} {\bibfnamefont
  {F.}~\bibnamefont {Cerutti}}, \bibinfo {author} {\bibfnamefont
  {A.}~\bibnamefont {Empl}}, \bibinfo {author} {\bibfnamefont {A.}~\bibnamefont
  {Fassò}}, \bibinfo {author} {\bibfnamefont {A.}~\bibnamefont {Ferrari}},
  \bibinfo {author} {\bibfnamefont {G.}~\bibnamefont {E.}}, \bibinfo {author}
  {\bibfnamefont {M.~V.}\ \bibnamefont {Garzelli}}, \bibinfo {author}
  {\bibfnamefont {K.}~\bibnamefont {Lee}}, \bibinfo {author} {\bibfnamefont
  {A.}~\bibnamefont {Ottolenghi}}, \bibinfo {author} {\bibfnamefont
  {M.}~\bibnamefont {Pelliccioni}}, \bibinfo {author} {\bibfnamefont {L.~S.}\
  \bibnamefont {Pinsky}}, \bibinfo {author} {\bibfnamefont {J.}~\bibnamefont
  {Ranft}}, \bibinfo {author} {\bibfnamefont {S.}~\bibnamefont {Roesler}},
  \bibinfo {author} {\bibfnamefont {P.~R.}\ \bibnamefont {Sala}},\ and\
  \bibinfo {author} {\bibfnamefont {T.~L.}\ \bibnamefont {Wilson}},\ }\bibfield
   {title} {\bibinfo {title} {{The fluka code for space applications: recent
  developments}},\ }\href
  {https://doi.org/https://doi.org/10.1016/j.asr.2003.03.045} {\bibfield
  {journal} {\bibinfo  {journal} {Adv. Space Res.}\ }\textbf {\bibinfo {volume}
  {34}},\ \bibinfo {pages} {1302} (\bibinfo {year} {2004})}\BibitemShut
  {NoStop}%
\bibitem [{\citenamefont {Roesler}\ \emph {et~al.}(2001)\citenamefont
  {Roesler}, \citenamefont {Engel},\ and\ \citenamefont
  {Ranft}}]{2001_RoeslerS}%
  \BibitemOpen
  \bibfield  {author} {\bibinfo {author} {\bibfnamefont {S.}~\bibnamefont
  {Roesler}}, \bibinfo {author} {\bibfnamefont {R.}~\bibnamefont {Engel}},\
  and\ \bibinfo {author} {\bibfnamefont {J.}~\bibnamefont {Ranft}},\ }\bibfield
   {title} {\bibinfo {title} {{The monte carlo event generator dpmjet-iii}},\
  }in\ \href@noop {} {\emph {\bibinfo {booktitle} {Advanced Monte Carlo for
  radiation physics, particle transport simulation and applications}}}\
  (\bibinfo  {publisher} {Springer},\ \bibinfo {year} {2001})\ pp.\ \bibinfo
  {pages} {1033--1038}\BibitemShut {NoStop}%
\bibitem [{\citenamefont {Nakamura}\ and\ \citenamefont
  {Hayakawa}(2015)}]{2015_NakamuraT}%
  \BibitemOpen
  \bibfield  {author} {\bibinfo {author} {\bibfnamefont {T.}~\bibnamefont
  {Nakamura}}\ and\ \bibinfo {author} {\bibfnamefont {T.}~\bibnamefont
  {Hayakawa}},\ }\bibfield  {title} {\bibinfo {title} {{Laser-driven γ-ray,
  positron, and neutron source from ultra-intense laser-matter interactions}},\
  }\href {https://doi.org/10.1063/1.4928889} {\bibfield  {journal} {\bibinfo
  {journal} {Phys. Plasmas}\ }\textbf {\bibinfo {volume} {22}},\ \bibinfo
  {pages} {083113} (\bibinfo {year} {2015})}\BibitemShut {NoStop}%
\bibitem [{\citenamefont {Köhn}\ and\ \citenamefont
  {Ebert}(2014)}]{2014_KoehnCh}%
  \BibitemOpen
  \bibfield  {author} {\bibinfo {author} {\bibfnamefont {C.}~\bibnamefont
  {Köhn}}\ and\ \bibinfo {author} {\bibfnamefont {U.}~\bibnamefont {Ebert}},\
  }\bibfield  {title} {\bibinfo {title} {{Angular distribution of
  Bremsstrahlung photons and of positrons for calculations of terrestrial
  gamma-ray flashes and positron beams}},\ }\href
  {https://doi.org/https://doi.org/10.1016/j.atmosres.2013.03.012} {\bibfield
  {journal} {\bibinfo  {journal} {Atmos Res}\ }\textbf {\bibinfo {volume}
  {135-136}},\ \bibinfo {pages} {432} (\bibinfo {year} {2014})}\BibitemShut
  {NoStop}%
\bibitem [{\citenamefont {Olsen}(1963)}]{1963_OlsenH}%
  \BibitemOpen
  \bibfield  {author} {\bibinfo {author} {\bibfnamefont {H.}~\bibnamefont
  {Olsen}},\ }\bibfield  {title} {\bibinfo {title} {{Opening angles of
  electron-positron pairs}},\ }\href {https://doi.org/10.1103/PhysRev.131.406}
  {\bibfield  {journal} {\bibinfo  {journal} {Phys. Rev.}\ }\textbf {\bibinfo
  {volume} {131}},\ \bibinfo {pages} {406} (\bibinfo {year}
  {1963})}\BibitemShut {NoStop}%
\bibitem [{\citenamefont {Azzam}\ \emph {et~al.}(2014)\citenamefont {Azzam},
  \citenamefont {Said},\ and\ \citenamefont {Al-abyad}}]{2014_AzzamA}%
  \BibitemOpen
  \bibfield  {author} {\bibinfo {author} {\bibfnamefont {A.}~\bibnamefont
  {Azzam}}, \bibinfo {author} {\bibfnamefont {S.~A.}\ \bibnamefont {Said}},\
  and\ \bibinfo {author} {\bibfnamefont {M.}~\bibnamefont {Al-abyad}},\
  }\bibfield  {title} {\bibinfo {title} {{Evaluation of different production
  routes for the radio medical isotope 203Pb using TALYS 1.4 and EMPIRE 3.1
  code calculations}},\ }\href
  {https://doi.org/https://doi.org/10.1016/j.apradiso.2014.05.009} {\bibfield
  {journal} {\bibinfo  {journal} {Appl Radiat Isot}\ }\textbf {\bibinfo
  {volume} {91}},\ \bibinfo {pages} {109} (\bibinfo {year} {2014})}\BibitemShut
  {NoStop}%
\bibitem [{\citenamefont {Tadamura}\ \emph {et~al.}(1999)\citenamefont
  {Tadamura}, \citenamefont {Kudoh}, \citenamefont {Motooka}, \citenamefont
  {Inubushi}, \citenamefont {Shirakawa}, \citenamefont {Hattori}, \citenamefont
  {Okada}, \citenamefont {Matsuda}, \citenamefont {Koshiji}, \citenamefont
  {Nishimura}, \citenamefont {Matsuda},\ and\ \citenamefont
  {Konishi}}]{1999_TadamuraE}%
  \BibitemOpen
  \bibfield  {author} {\bibinfo {author} {\bibfnamefont {E.}~\bibnamefont
  {Tadamura}}, \bibinfo {author} {\bibfnamefont {T.}~\bibnamefont {Kudoh}},
  \bibinfo {author} {\bibfnamefont {M.}~\bibnamefont {Motooka}}, \bibinfo
  {author} {\bibfnamefont {M.}~\bibnamefont {Inubushi}}, \bibinfo {author}
  {\bibfnamefont {S.}~\bibnamefont {Shirakawa}}, \bibinfo {author}
  {\bibfnamefont {N.}~\bibnamefont {Hattori}}, \bibinfo {author} {\bibfnamefont
  {T.}~\bibnamefont {Okada}}, \bibinfo {author} {\bibfnamefont
  {T.}~\bibnamefont {Matsuda}}, \bibinfo {author} {\bibfnamefont
  {T.}~\bibnamefont {Koshiji}}, \bibinfo {author} {\bibfnamefont
  {K.}~\bibnamefont {Nishimura}}, \bibinfo {author} {\bibfnamefont
  {K.}~\bibnamefont {Matsuda}},\ and\ \bibinfo {author} {\bibfnamefont
  {J.}~\bibnamefont {Konishi}},\ }\bibfield  {title} {\bibinfo {title}
  {{Assessment of regional and global left ventricular function by reinjection
  Tl-201 and rest Tc-99m sestamibi ECG-gated SPECT}},\ }\href
  {https://doi.org/10.1016/S0735-1097(98)00661-5} {\bibfield  {journal}
  {\bibinfo  {journal} {J. Am. Coll. Cardiol.}\ }\textbf {\bibinfo {volume}
  {33}},\ \bibinfo {pages} {991} (\bibinfo {year} {1999})}\BibitemShut
  {NoStop}%
\end{thebibliography}%

\end{document}